\PassOptionsToPackage{unicode}{hyperref}
\PassOptionsToPackage{hyphens}{url}
\PassOptionsToPackage{dvipsnames,svgnames,x11names}{xcolor}
\documentclass[
  12pt]{article}

\usepackage{amsmath,amssymb}
\usepackage{iftex}
\ifPDFTeX
  \usepackage[T1]{fontenc}
  \usepackage[utf8]{inputenc}
  \usepackage{textcomp} 
\else 
  \usepackage{unicode-math}
  \defaultfontfeatures{Scale=MatchLowercase}
  \defaultfontfeatures[\rmfamily]{Ligatures=TeX,Scale=1}
\fi
\usepackage{lmodern}
\ifPDFTeX\else  
\fi
\IfFileExists{upquote.sty}{\usepackage{upquote}}{}
\IfFileExists{microtype.sty}{
  \usepackage[]{microtype}
  \UseMicrotypeSet[protrusion]{basicmath} 
}{}
\makeatletter
\@ifundefined{KOMAClassName}{
  \IfFileExists{parskip.sty}{%
    \usepackage{parskip}
  }{
    \setlength{\parindent}{0pt}
    \setlength{\parskip}{6pt plus 2pt minus 1pt}}
}{
  \KOMAoptions{parskip=half}}
\makeatother
\usepackage{xcolor}
\makeatletter
\ifx\paragraph\undefined\else
  \let\oldparagraph\paragraph
  \renewcommand{\paragraph}{
    \@ifstar
      \xxxParagraphStar
      \xxxParagraphNoStar
  }
  \newcommand{\xxxParagraphStar}[1]{\oldparagraph*{#1}\mbox{}}
  \newcommand{\xxxParagraphNoStar}[1]{\oldparagraph{#1}\mbox{}}
\fi
\ifx\subparagraph\undefined\else
  \let\oldsubparagraph\subparagraph
  \renewcommand{\subparagraph}{
    \@ifstar
      \xxxSubParagraphStar
      \xxxSubParagraphNoStar
  }
  \newcommand{\xxxSubParagraphStar}[1]{\oldsubparagraph*{#1}\mbox{}}
  \newcommand{\xxxSubParagraphNoStar}[1]{\oldsubparagraph{#1}\mbox{}}
\fi
\makeatother

\usepackage{longtable,booktabs,array}
\usepackage{calc} 
\usepackage{etoolbox}
\makeatletter
\patchcmd\longtable{\par}{\if@noskipsec\mbox{}\fi\par}{}{}
\makeatother
\IfFileExists{footnotehyper.sty}{\usepackage{footnotehyper}}{\usepackage{footnote}}
\makesavenoteenv{longtable}
\usepackage{graphicx}
\makeatletter
\def\maxwidth{\ifdim\Gin@nat@width>\linewidth\linewidth\else\Gin@nat@width\fi}
\def\maxheight{\ifdim\Gin@nat@height>\textheight\textheight\else\Gin@nat@height\fi}
\makeatother
\setkeys{Gin}{width=\maxwidth,height=\maxheight,keepaspectratio}
\makeatletter
\def\fps@figure{htbp}
\makeatother

\makeatletter
\@ifpackageloaded{caption}{}{\usepackage{caption}}
\AtBeginDocument{%
\ifdefined\contentsname
  \renewcommand*\contentsname{Table of contents}
\else
  \newcommand\contentsname{Table of contents}
\fi
\ifdefined\listfigurename
  \renewcommand*\listfigurename{List of Figures}
\else
  \newcommand\listfigurename{List of Figures}
\fi
\ifdefined\listtablename
  \renewcommand*\listtablename{List of Tables}
\else
  \newcommand\listtablename{List of Tables}
\fi
\ifdefined\figurename
  \renewcommand*\figurename{Figure}
\else
  \newcommand\figurename{Figure}
\fi
\ifdefined\tablename
  \renewcommand*\tablename{Table}
\else
  \newcommand\tablename{Table}
\fi
}
\@ifpackageloaded{float}{}{\usepackage{float}}
\floatstyle{ruled}
\@ifundefined{c@chapter}{\newfloat{codelisting}{h}{lop}}{\newfloat{codelisting}{h}{lop}[chapter]}
\floatname{codelisting}{Listing}

\makeatother
\makeatletter
\@ifpackageloaded{caption}{}{\usepackage{caption}}
\@ifpackageloaded{subcaption}{}{\usepackage{subcaption}}
\makeatother

\ifLuaTeX
  \usepackage{selnolig}  
\fi
\usepackage[]{natbib}
\usepackage{bookmark}

\IfFileExists{xurl.sty}{\usepackage{xurl}}{} 
\hypersetup{
  pdftitle={Title},
  pdfauthor={Author 1; Author 2},
  pdfkeywords={3 to 6 keywords, that do not appear in the title},
  colorlinks=true,
  linkcolor={blue},
  filecolor={Maroon},
  citecolor={Blue},
  urlcolor={Blue},
  pdfcreator={LaTeX via pandoc}}

\newcommand{\anon}{1}

\begin{document}

\def\spacingset#1{\renewcommand{\baselinestretch}%
{#1}\small\normalsize} \spacingset{1}


\if1\anon
{
  \title{\bfseries Fast simulation of Gaussian random fields with flexible correlation models in Euclidean spaces}
  \author{
    Moreno Bevilacqua\thanks{
      Facultad de Ingenier\'ia y Ciencias, Universidad Adolfo Ib\'a\~nez, Chile;
      Dipartimento di Scienze Ambientali, Informatica e Statistica, Universit\`a Ca' Foscari, Italy.
    }
    \and
    Xavier Emery\thanks{
      Department of Mining Engineering, Universidad de Chile, Santiago, Chile;
      Advanced Mining Technology Center, Universidad de Chile, Santiago, Chile.
    }
    \and
    Francisco Cuevas-Pacheco\thanks{
      Departamento de Matem\'atica, Universidad T\'ecnica Federico Santa Mar\'ia, Chile.
    }
  }
  \maketitle
} \fi

\if0\anon
{
  \bigskip
  \bigskip
  \bigskip
  \begin{center}
    {\LARGE\bf Fast simulation of Gaussian random fields with flexible correlation models in Euclidean spaces}
\end{center}
  
} \fi

\bigskip
\begin{abstract}
The efficient simulation of Gaussian random fields with flexible correlation structures is fundamental in spatial statistics, machine learning, and uncertainty quantification. In this work, we revisit the spectral turning-bands (STB) method as a versatile and scalable framework for simulating isotropic Gaussian random fields with a broad range of covariance models. Beyond the Matérn model, we extend the STB approach to the Kummer--Tricomi model, which exhibits polynomially decaying correlations and long-range dependence, and to the Gauss--Hypergeometric model, which yields compactly supported correlations, including the generalized Wendland class. 
For both classes, we derive exact sampling schemes for the spectral variables entering the STB construction.
The resulting simulator yields the usual finite-$L$ STB approximation to the target Gaussian random field while remaining numerically stable and computationally efficient, with computational complexity linear in the number of locations and spectral components.
Numerical experiments confirm good accuracy and stable performance across a broad range of parameter settings. Supplementary materials for this article are available online.
\end{abstract}

\noindent%

\vfill

\newpage
\spacingset{1.8} 

\section{Introduction}

\subsection{Motivation}

Gaussian random fields (RFs) indexed over Euclidean spaces play a central role in spatial statistics, geostatistics, machine learning, numerical analysis, design of experiments, signal processing, and uncertainty quantification, among many other areas
\citep{Cressie:1993,stein-book,Banerjee-Carlin-Gelfand:2004,williams2006gaussian,wendland2004scattered,cp2003}.
A Gaussian RF is fully characterized by its mean function and covariance structure, the latter governing both smoothness and spatial dependence.

In many applications, from spatial interpolation to Bayesian inverse problems, fast simulation under a prescribed covariance model is essential for both inference and uncertainty propagation through complex models.
Fast Gaussian RF simulation is also important in likelihood-free inference, including Approximate Bayesian Computation \citep{sisson2018handbook} and simulation-based amortized methods \citep{zammit2025neural}, as well as in simulation-based hypothesis testing, for example in the Global Rank Envelope Test \citep{globale}.
It also plays an important role in numerical weather prediction and ensemble forecasting, where spatially coherent error fields are used to generate ensembles of weather fields for postprocessing and uncertainty quantification \citep{BerrocalRafteryGneiting2007}.

Simulation is equally important for non-Gaussian RFs obtained from latent Gaussian structures.
Relevant examples include Gaussian copula models \citep{MasarottoVarin2012}, truncated Gaussian and plurigaussian RFs \citep{armstrong2011plurigaussian}, substitution RFs \citep{emery2008substitution}, chi-squared and non-central chi-squared RFs \citep{Adler:1981,emery2007}, Tukey--$g$--and--$h$ RFs \citep{XuGenton2017}, Weibull RFs \citep{Bevilacqua2020}, and Poisson-based RFs \citep{MoralesNavarrete2024}.
In all these settings, efficient simulation of the underlying Gaussian field is a key computational ingredient.

\subsection{Review of existing methods}

One of the earliest and most widely used methods for simulating a Gaussian RF with a prescribed covariance function at a finite set of locations is the \emph{Cholesky method} \citep{Ripley:1987}.
It relies on the Cholesky factorization of the covariance matrix and has computational cost $O(N^3)$, where $N$ denotes the number of locations.
This quickly becomes prohibitive for large datasets.

To reduce this burden, several alternative approaches exploit structure in the covariance or precision matrix.
For regular grids, the \emph{circulant embedding} method \citep{Dietrich-Newsam:1993,chan1997algorithm} uses the Fast Fourier Transform (FFT) after embedding the grid into a torus, yielding computational cost of order $O(N\log N)$.
Its applicability is however limited to regular lattices in Euclidean spaces, and additional corrections may be required when the embedded covariance matrix fails to be nonnegative definite \citep{Chiles2012}.

For irregular spatial designs, an important class of methods is based on the \emph{Vecchia approximation} \citep{vecchia1988} and its extensions \citep{datta2016a,katzfuss2017general}.
These approaches exploit approximate conditional independence through sparse factorizations of the joint density or precision structure.
Conditioning each location on a small set of neighbors, say $m\ll N$, yields sequential simulation with cost of order $O(Nm^2)$.
Accuracy depends strongly on both the ordering of the locations and the choice of neighbors.
Vecchia-type methods are often effective for covariance models with localized dependence in the precision structure, such as Mat\'ern models, although their behavior can be less favorable for compactly supported covariance models, where the precision matrix need not inherit the sparsity of the covariance matrix \citep{mateporcu}.

For the Mat\'ern class, an alternative route is provided by the \emph{SPDE--INLA} framework \citep{Lindgren:Rue:Lindstrom:2011}, in which the continuous field is approximated on a finite-element mesh by a Gaussian Markov RF with sparse precision matrix.
This allows simulation through sparse linear algebra rather than dense covariance factorizations, typically with computational complexity around $O(N^{3/2})$ in two dimensions, and sometimes closer to linear complexity depending on the mesh structure and implementation.

General-purpose simulation can also be performed via Markov chain Monte Carlo (MCMC) schemes such as Gibbs sampling \citep{Geman:Geman:1984} and Metropolis--Hastings \citep{Hastings:1970}.
These methods can accommodate irregular sampling configurations, but they are usually not competitive for large-scale simulation because they require many iterations and repeated sampling from conditional Gaussian distributions.

For compactly supported covariance functions, sparse Cholesky factorization of the covariance matrix can also be effective \citep{da2006,da1989}, provided that the covariance matrix is sufficiently sparse.
In practice, the resulting gains may be substantial only when the support is small relative to the spatial domain and the sampling design.

A conceptually different strategy for general stationary covariance models is the continuous spectral method of \citet{Shinozuka1971,Shinozuka1972}, which applies whenever the spectral density is available.
This approach can be interpreted as a particular case of the more general \emph{turning-bands} method introduced by \citet{Matheron:1973}, where one-dimensional processes are spread over a multidimensional Euclidean space.
In the spectral setting, the one-dimensional components reduce to cosine waves, leading to what we call the \emph{spectral turning--bands} (STB) method, following \citet{Mantoglou:1987} and \citet{emery2016}.

The STB method is attractive because it reduces simulation to the generation of univariate components along randomly oriented lines, while remaining applicable to arbitrary spatial designs.
For a finite number of spectral components, say $L$, the resulting field is a finite-$L$ STB approximation to the target Gaussian RF.
Increasing $L$ improves the approximation through a central-limit effect, so that $L$ controls the trade-off between computational cost and fidelity to the target field.
The overall computational cost scales linearly in both the number of locations and the number of spectral components, namely $O(NL)$.

\subsection{A brief computational comparison}

To illustrate the computational performance of the proposed STB approach, we compare its runtime with two widely used methods for large-scale simulation on irregular spatial designs: the SPDE--INLA framework \citep{Lindgren:Rue:Lindstrom:2011} and the Vecchia approximation \citep{vecchia1988}, implemented in the \texttt{INLA} and \texttt{GpGp} \texttt{R} packages, respectively.
Runtime is measured in terms of elapsed  time using the \texttt{system.time} function in \texttt{R}. Although the main contribution of this paper is the extension of STB simulation beyond the Mat\'ern class, we focus this benchmark on the Mat\'ern model, for which both SPDE--INLA and Vecchia are well-established reference methods. All computations were performed on a 2$\times$8-core CPU (Intel Xeon E5-2689 @ 1200--3600 MHz) with 128 GB RAM.

In all experiments, we consider the Mat\'ern covariance model $\mathcal{M}_{\nu,\alpha}$ with marginal variance $\sigma^2=2$, scale parameter $\alpha=0.12$, and two smoothness settings, namely $\nu=0.5$ and $\nu=1$.
Sampling locations are generated independently from the uniform distribution on $[0,1]^2$, and the sample size ranges from $n=100$ to $n=100{,}000$.

For the Vecchia approximation, we consider conditioning set sizes $m\in\{5,10,20,30\}$ and use the function \texttt{fast\_Gp\_sim}.
For SPDE--INLA, we construct a constrained Delaunay triangulation over $[0,1]^2$ using \texttt{fm\_mesh\_2d\_inla} from the \texttt{fmesher} package, with inner and outer offsets equal to $0.1$ and maximum edge length equal to $0.072$ in both regions.
The Mat\'ern SPDE model is specified through \texttt{inla.spde2.matern}, with $\alpha=\nu+d/2$ \citep{Lindgren:Rue:Lindstrom:2011}, and with the log-linear parametrization
\[
\log\kappa_0 = \tfrac{1}{2}\log(8\nu) - \log(r_0),
\qquad
\log\tau_0 = \tfrac{1}{2}
\bigl[\log\Gamma(\nu)-\log\Gamma(\alpha)-\log(4\pi)
-2\nu\log\kappa_0-\log\sigma^2\bigr],
\]
where $r_0=0.12$ and $\sigma^2=2$, so that the target practical range and marginal variance are aligned with those used by the competing methods.
Weakly informative Gaussian priors with mean zero and precision $0.1$ are assigned to both $\log\tau$ and $\log\kappa$.
The precision matrix $\mathbf{Q}$ is computed using \texttt{inla.spde.precision} at $\boldsymbol{\theta}=(0,0)^\top$.
For INLA, we report separately the one-time cost of mesh construction and precision-matrix assembly, and the per-realization simulation time obtained through \texttt{inla.qsample}; only the latter is included in the replication-wise timing comparison.
For each configuration, runtimes are summarized over $n_{\rm sim}=1{,}000$ independent realizations, computed in parallel using \texttt{doParallel} on half of the available cores.

\begin{figure}[htbp]
    \centering
    \includegraphics[width=\textwidth]{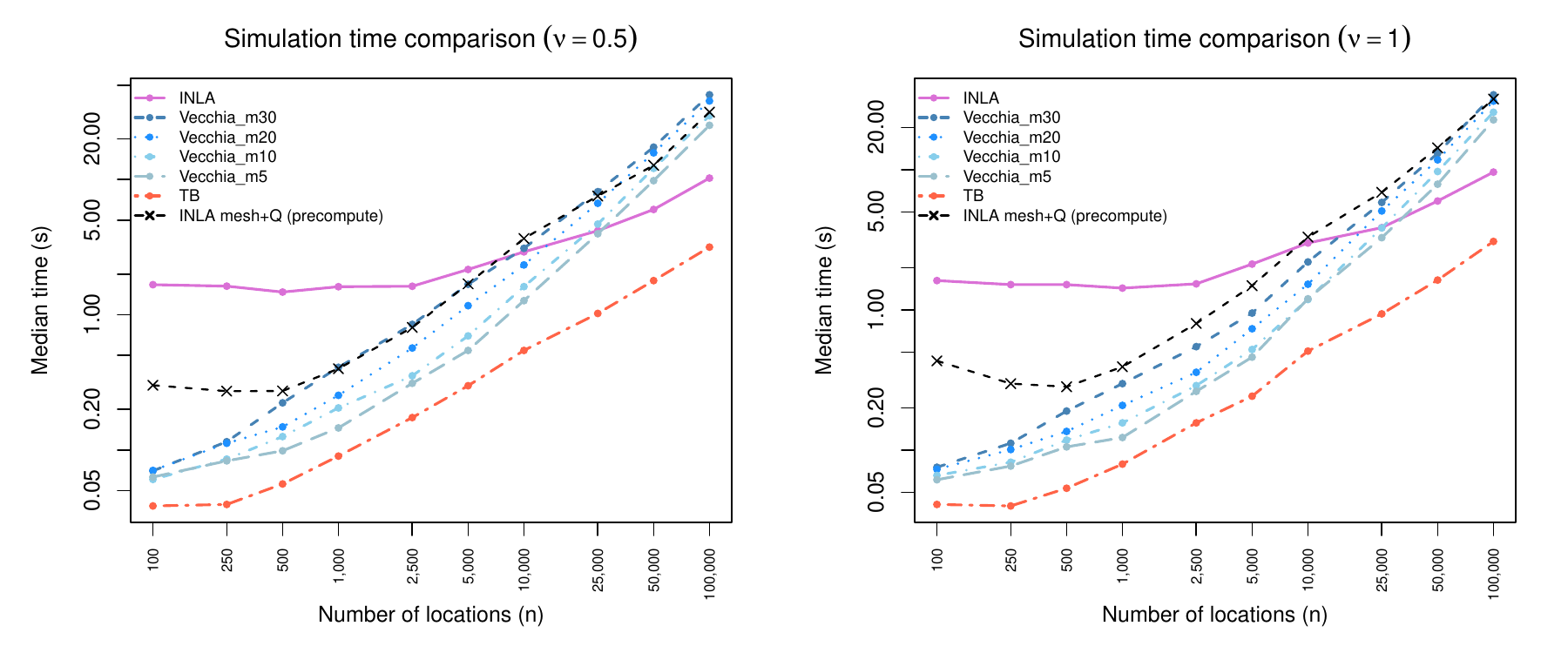}
    \caption{Median simulation time (in seconds) versus the number of spatial locations $n$, shown on a log--log scale, for SPDE--INLA sampling, Vecchia approximations with $m=5,10,20,30$, and the proposed STB algorithm. The left panel corresponds to $\nu=0.5$, and the right panel to $\nu=1$ . Black crosses connected by a dashed line indicate the one-time cost of mesh construction and precision-matrix assembly for INLA; this cost is not included in the per-realization simulation times.}
    \label{fig:timing}
\end{figure}

The results are reported in Figure~\ref{fig:timing}.
For small sample sizes ($n\le 1{,}000$), all methods are computationally inexpensive and runtime differences remain modest.
As $n$ increases, the proposed STB simulator exhibits the most favorable scaling behavior under both smoothness settings.
For example, when $n=100{,}000$ and $\nu=0.5$, the median simulation time for STB is \texttt{3.1650} seconds, compared with \texttt{25.204}, \texttt{29.656}, \texttt{38.164}, and \texttt{42.385} seconds for the Vecchia approximation with $m=5,10,20,30$, respectively, and \texttt{10.257} seconds for INLA sampling alone.
The latter does not include the additional one-time mesh and precision-matrix construction cost of \texttt{31.57} seconds.
A similar pattern is observed for $\nu=1$, where the corresponding median times at $n=100{,}000$ are \texttt{3.077} seconds for STB and \texttt{9.605} seconds for INLA.
Overall, these results are consistent with the near-linear scaling of STB in the number of locations, as predicted by its $O(NL)$ complexity, and motivate its use for large-scale simulation on irregular designs.

Table~\ref{tab:sim-comparison} summarizes the main computational and qualitative features of the simulation methods discussed above.
For irregularly distributed locations, STB is particularly attractive among the methods considered here because it combines linear complexity in the number of locations with modest memory requirements, while avoiding both dense matrix factorizations and mesh construction.

\begin{table}[htbp]
\centering
\caption{Computational and qualitative comparison of simulation methods for Gaussian RFs. Here $N$ denotes the number of locations, $m$ the Vecchia conditioning size, and $L$ the number of spectral components in STB.}
\label{tab:sim-comparison}
\begin{tabular}{lcccc}
\hline
Method & Time & Memory & Design & Exactness \\
\hline
Cholesky & $O(N^{3})$ & $O(N^{2})$ & arbitrary & exact \\
Circulant embedding & $O(N\log N)$ & $O(N)$ & regular grid & exact$^{\dagger}$ \\
Vecchia & $O(Nm^{2})$ & $O(Nm)$ & arbitrary & approx.$^{\ddagger}$ \\
SPDE--INLA & $\approx O(N^{3/2})$ (2D) & $O(N)$ (sparse) & mesh & approx.$^{\S}$ \\
STB & $O(NL)$ & $O(N{+}L)$ & arbitrary & finite-$L$ approx.$^{\P}$ \\
\hline
\end{tabular}

\vspace{2mm}
{\footnotesize
$^{\dagger}$ Exact when the circulant embedding is valid, i.e., when the embedded covariance matrix is nonnegative definite.
$^{\ddagger}$ Exact for the approximating Vecchia model.
$^{\S}$ Exact sampling from the discretized GMRF defined on the mesh.
$^{\P}$ Exact spectral-frequency sampling; Gaussianity is recovered as $L\to\infty$.
}
\end{table}

\subsection{Contribution}

This work develops STB simulation methods for Gaussian RFs in Euclidean spaces under flexible isotropic covariance models.
Here, flexibility refers to parametric covariance families that provide continuous control over the smoothness of the field, while also allowing either compact support or heavier-than-Mat\'ern tail behavior.

The Mat\'ern ($\mathcal{M}$) family \citep{Stein:1999,mateporcu} is the benchmark model in spatial statistics because of its interpretable parameters and explicit control of mean-square differentiability.
However, as discussed in \citet{mateporcu}, it has two well-known limitations:
global support and exponentially decaying correlations.

Two recent covariance families overcome these limitations in complementary directions:
\begin{enumerate}
\item The \emph{Gauss--Hypergeometric} ($\mathcal{GH}$) model \citep{emery2022gauss,hyp2025}, which defines a broad class of compactly supported correlations including, as special cases, the generalized Wendland family \citep{gnei02,BFFP}.
It preserves the smoothness parametrization of the Mat\'ern model while introducing an additional shape parameter.

\item The \emph{Kummer--Tricomi} ($\mathcal{K}$) model \citep{ma2022beyond}, which introduces an additional tail parameter allowing polynomial decay while maintaining smoothness flexibility in all spatial dimensions.
\end{enumerate}

Both models can therefore be viewed as extensions of the Mat\'ern class:
the $\mathcal{GH}$ model toward compactly supported correlations, and the $\mathcal{K}$ model toward heavy-tailed and long-range dependence.

While STB simulation for the Mat\'ern model is classical and well understood \citep{EmeryLantu2006}, extending the approach to the $\mathcal{K}$ and $\mathcal{GH}$ families requires new stochastic representations for their spectral distributions.
The main difficulty is that the corresponding spectral densities involve special hypergeometric functions and do not appear to admit simple direct samplers, except in a few special cases \citep[e.g.,][]{arroyo2021}.
To address this issue, we derive exact sampling representations for the spectral variables entering the STB construction, together with corresponding stochastic samplers that avoid numerical inversion, quadrature, or other approximations at the spectral-sampling stage.
Accordingly, exactness in this paper refers to exact sampling of the spectral variables entering the STB construction; the resulting RF remains the usual finite-$L$ STB approximation to the target Gaussian field.

More specifically:
\begin{itemize}
\item we show that the $\mathcal{K}$ model admits a tractable spectral representation leading to exact sampling of the STB spectral variable through a Beta--prime Gaussian mixture;
\item we show that STB simulation for the $\mathcal{GH}$ class can be developed through two alternative spectral constructions based on Beta and Gasper mixtures.
\end{itemize}

The resulting algorithms provide a unified stochastic framework for fast simulation of Gaussian RFs with $\mathcal{M}$, $\mathcal{K}$, and $\mathcal{GH}$ covariance functions over very large sets of spatial locations.
They are implemented in the function \texttt{GeoSimapprox} of the \texttt{R} package \texttt{GeoModels} \citep{geomodels}, which supports simulation of Gaussian and a broad class of non-Gaussian spatial fields with $\mathcal{M}$, $\mathcal{K}$, and $\mathcal{GH}$ underlying correlation models.

Two numerical studies illustrate the proposed methodology.
First, we conduct a simulation study assessing the accuracy of the STB algorithms for the $\mathcal{GH}$ and $\mathcal{K}$ models across a broad range of parameter settings by comparing empirical semivariograms with their theoretical counterparts.
Second, Section~B of the Supplementary Material analyzes a large climate dataset and shows how the proposed STB simulators can be used for parametric bootstrap in the computation of standard errors and model selection criteria under weighted pairwise composite likelihood \citep{NNP}, for both Gaussian and transformed Gaussian RFs with $\mathcal{M}$, $\mathcal{K}$, and generalized Wendland correlation functions.

\subsection{Outline}

The remainder of the paper is organized as follows.
Section~\ref{sec2} recalls the spectral representation of stationary Gaussian RFs and the STB framework.
Section~\ref{sec3} reviews the $\mathcal{M}$ correlation model and its STB sampler.
Section~\ref{sec4} develops the $\mathcal{K}$ model, its spectral representation, and the associated STB algorithm.
Section~\ref{sec5} introduces the $\mathcal{GH}$ family, including the Beta- and Gasper-mixture formulations and the corresponding STB algorithms.
Section~\ref{sec6} reports numerical experiments assessing the accuracy and numerical stability of the proposed samplers.
Section~\ref{sec8} concludes with a discussion and directions for future work.

\section{Spectral representation and spectral turning--bands simulation}
\label{sec2}

Let $\{Z(\mathbf{s}) : \mathbf{s}\in D\}$ be a real-valued, zero-mean stationary Gaussian RF defined on a bounded domain
$D\subset\mathbb{R}^d$, with stationary covariance
\[
C(\mathbf{h})=\mathrm{Cov}\!\bigl(Z(\mathbf{s}),Z(\mathbf{s}+\mathbf{h})\bigr),
\qquad
\mathbf{h}\in\mathbb{R}^d,
\]
and variance $\sigma^2:=C(\mathbf{0})$.
The corresponding correlation function is
\[
\phi(\mathbf{h})=\frac{C(\mathbf{h})}{\sigma^2},
\qquad
\phi(\mathbf{0})=1.
\]

By Bochner's theorem, any continuous positive-semidefinite function $C:\mathbb{R}^d\to\mathbb{C}$ admits the representation \citep{Yaglom:1987}
\begin{equation}\label{eq:bochner}
C(\mathbf{h})=\int_{\mathbb{R}^d} e^{\mathsf{i}\boldsymbol{\omega}\cdot\mathbf{h}}\,F({\rm d}\boldsymbol{\omega}),
\end{equation}
where $\mathsf{i}$ is the imaginary unit, $\cdot$ is the Euclidean scalar product, and $F$ is a finite nonnegative symmetric measure on $\mathbb{R}^d$.
If $F$ admits a density $f_C$, then
\[
C(\mathbf{h})=\int_{\mathbb{R}^d} e^{\mathsf{i}\boldsymbol{\omega}\cdot\mathbf{h}} f_C(\boldsymbol{\omega})\,{\rm d}\boldsymbol{\omega},
\qquad
f_C(\boldsymbol{\omega})\ge 0,
\qquad
\int_{\mathbb{R}^d}f_C(\boldsymbol{\omega})\,{\rm d}\boldsymbol{\omega}=\sigma^2.
\]
Since the field is real-valued, $C(\mathbf{h})=C(-\mathbf{h})$, and therefore
\[
C(\mathbf{h})=\int_{\mathbb{R}^d}\cos(\boldsymbol{\omega}\cdot\mathbf{h})\,f_C(\boldsymbol{\omega})\,{\rm d}\boldsymbol{\omega}.
\]

When the spectral measure is absolutely continuous, we denote by $f$ the normalized spectral density of the correlation function $\phi$, namely
\[
f(\boldsymbol{\omega})=\frac{1}{\sigma^2}f_C(\boldsymbol{\omega}),
\qquad
\int_{\mathbb{R}^d}f(\boldsymbol{\omega})\,{\rm d}\boldsymbol{\omega}=1,
\]
so that
\begin{equation}\label{eq:spectral-corr}
\phi(\mathbf{h})
=\int_{\mathbb{R}^d}\cos(\boldsymbol{\omega}\cdot\mathbf{h})\,f(\boldsymbol{\omega})\,{\rm d}\boldsymbol{\omega}.
\end{equation}

For an isotropic RF, $\phi(\mathbf{h})=\phi(x)$ with $x=\|\mathbf{h}\|$, and, in the absolutely continuous case, the spectral density depends only on $r=\|\boldsymbol{\omega}\|$, that is,
$
f(\boldsymbol{\omega})=g_R(r)
$
for some radial function $g_R$.
Integrating~\eqref{eq:spectral-corr} over directions yields the Schoenberg representation \citep{Schoenberg1938b}
\begin{equation}\label{eq:schoenberg}
\phi(x)=\int_0^\infty \Omega_d(xr)\,|\mathbb{S}^{d-1}|\,r^{d-1}\,g_R(r)\,{\rm d}r,
\qquad x\ge 0,
\end{equation}
where $|\mathbb{S}^{d-1}|=\frac{2\pi^{d/2}}{\Gamma(d/2)}$ is the surface area of the unit sphere in $\mathbb{R}^d$, and
\[
\Omega_d(x)=\Gamma\!\Bigl(\tfrac{d}{2}\Bigr)\Bigl(\tfrac{2}{x}\Bigr)^{d/2-1}J_{d/2-1}(x),
\qquad x>0,
\]
with $J_{\nu}$ denoting the Bessel function of the first kind of order $\nu$.
We denote by $\Phi_d$ the class of continuous isotropic positive-definite functions on $\mathbb{R}^d$ satisfying $\phi(0)=1$; these classes are nested as
\[
\Phi_1 \supset \Phi_2 \supset \cdots \supset \Phi_\infty := \bigcap_{d\ge1}\Phi_d.
\]

In the absolutely continuous isotropic case, let $\boldsymbol{\Omega}\in\mathbb{R}^d$ be a random frequency vector with density
$
f(\boldsymbol{\omega})=g_R(\|\boldsymbol{\omega}\|),
$
normalized so that
\[
\int_{\mathbb{R}^d}f(\boldsymbol{\omega})\,{\rm d}\boldsymbol{\omega}
=
|\mathbb{S}^{d-1}|\int_0^\infty r^{d-1}g_R(r)\,{\rm d}r
=1.
\]
Let $R=\|\boldsymbol{\Omega}\|$ denote the associated spectral radius.
Its density is
\begin{equation}\label{eq:fR}
f_R(r)=|\mathbb{S}^{d-1}|\,r^{d-1}\,g_R(r),
\qquad r\ge 0.
\end{equation}
With this normalization,
\begin{equation}\label{eq:spectral-expectation}
\phi(\|\mathbf{h}\|)
=
\mathbb{E}\!\left[\cos(\boldsymbol{\Omega}\cdot\mathbf{h})\right],
\qquad \mathbf{h}\in\mathbb{R}^d.
\end{equation}
Moreover, writing
\[
\boldsymbol{\Omega}=R\boldsymbol{\Theta},
\qquad
\boldsymbol{\Theta}\sim\mathrm{Unif}(\mathbb{S}^{d-1}),
\qquad
R\perp\!\!\!\perp\boldsymbol{\Theta},
\]
separates the radial and angular components of the spectral distribution and provides the basis for model-specific samplers.

The spectral turning--bands (STB) method exploits~\eqref{eq:spectral-expectation} by replacing the expectation with a Monte Carlo average over independent spectral components.
For a target covariance model $\sigma^2\phi$, the STB field is defined as \citep{Lantu2002}
\begin{equation}\label{eq:TB-def-prob}
\widetilde Z_L(\mathbf{s})
=
\sigma \sum_{\ell=1}^L
\sqrt{\frac{-2\ln(\varepsilon_\ell)}{L}}\,
\cos\!\bigl(\boldsymbol{\Omega}_\ell\cdot\mathbf{s}+\Phi_\ell\bigr),
\qquad \mathbf{s}\in D,
\end{equation}
where
\[
\varepsilon_\ell \stackrel{\text{i.i.d.}}{\sim}\mathrm{Unif}(0,1),
\qquad
\boldsymbol{\Omega}_\ell \stackrel{\text{i.i.d.}}{\sim} F,
\qquad
\Phi_\ell \stackrel{\text{i.i.d.}}{\sim}\mathrm{Unif}(0,2\pi),
\qquad
\ell=1,\dots,L,
\]
and $F$ denotes the spectral distribution of the target correlation model (with density $f$ whenever the spectral measure is absolutely continuous).

Following \cite{AELL}, the variable $\varepsilon_\ell$ is introduced so that, for each $\ell$ and each location $\mathbf{s}$, the random quantity
\[
\sqrt{-2\ln(\varepsilon_\ell)}\,
\cos\!\bigl(\boldsymbol{\Omega}_\ell\cdot\mathbf{s}+\Phi_\ell\bigr)
\]
has a standard Gaussian distribution \citep{box:1958}.
Consequently, $\widetilde Z_L(\mathbf{s})$ is Gaussian with mean zero and variance $\sigma^2$ at every fixed location $\mathbf{s}$.

By construction,
$
\mathbb{E}\!\bigl[\widetilde Z_L(\mathbf{s})\bigr]=0$,
$\mathbf{s}\in D,$
and, for any $\mathbf{s},\mathbf{t}\in D$,
$
\mathbb{E}\!\bigl[\widetilde Z_L(\mathbf{s})\widetilde Z_L(\mathbf{t})\bigr]
=
\sigma^2\,\phi(\|\mathbf{s}-\mathbf{t}\|).
$
Thus, for every finite $L$, the STB construction reproduces the target covariance exactly.
However, the field $\widetilde Z_L$ is generally not Gaussian as a process for finite $L$; only its one-dimensional marginals are exactly Gaussian.
The target Gaussian dependence structure is recovered only asymptotically: as $L\to\infty$, $\widetilde Z_L$ converges in finite-dimensional distributions to the target Gaussian RF by a central-limit effect, and the Monte Carlo error associated with the finite-$L$ approximation decreases at the usual rate $O(L^{-1/2})$.

When the spectral measure is absolutely continuous, isotropic spectral sampling reduces to sampling $R$ from~\eqref{eq:fR}, sampling $\boldsymbol{\Theta}\sim\mathrm{Unif}(\mathbb{S}^{d-1})$, and setting $\boldsymbol{\Omega}=R\boldsymbol{\Theta}$.
More generally, the key algorithmic step is to construct the model-specific spectral random element entering the STB representation.

Sections~\ref{sec3}--\ref{sec5} derive stochastic representations that allow exact sampling of the spectral variables for the Mat\'ern, Kummer--Tricomi, and Gauss--Hypergeometric models.
Accordingly, throughout the paper, exactness refers to the spectral-sampling step; the resulting RF remains the usual finite-$L$ STB approximation to the target Gaussian field.

Given $\phi\in\Phi_d$ with spectral distribution $F$ and sill $\sigma^2$, the generic STB algorithm is:
\begin{enumerate}
\item Choose the number of spectral components $L$.
\item For each $\ell=1,\dots,L$:
\begin{enumerate}
\item Sample the model-specific spectral variable(s) defining $\boldsymbol{\Omega}_\ell\sim F$;
\item Sample $\Phi_\ell\sim\mathrm{Unif}(0,2\pi)$;
\item Sample $\varepsilon_\ell\sim\mathrm{Unif}(0,1)$.
\end{enumerate}
\item Compute $\widetilde Z_L(\mathbf{s})$ via~\eqref{eq:TB-def-prob} for all $\mathbf{s}\in D$.
\end{enumerate}

The computational cost scales linearly in the number of spectral components and in the number of spatial locations at which the field is evaluated.
The resulting algorithm is straightforward to parallelize across frequencies and locations.
In practice, once sampling from the model-specific spectral representation is available, the additional computational overhead of the STB construction is modest, so that reasonably large values of $L$ can be used in applications.

\section{STB simulation for the Mat\'ern correlation model}
\label{sec3}

In this section, we review the Mat\'ern class $\mathcal{M}$ \citep{Stein:1999,mateporcu} and discuss its STB-based simulation.
The Mat\'ern class is one of the most widely used families of isotropic correlation functions, as it provides a simple and interpretable control over both smoothness and spatial range.
For parameters $\nu>0$ (smoothness) and $\alpha>0$ (scale), we adopt the standard parameterization \citep{Matern}:
\begin{equation}\label{eq:matern}
\mathcal{M}_{\nu,\alpha}(x)
= \frac{2^{1-\nu}}{\Gamma(\nu)}
\left(\frac{x}{\alpha}\right)^{\nu}
K_{\nu}\!\left(\frac{x}{\alpha}\right),
\qquad x \ge 0,
\end{equation}
where $K_{\nu}$ denotes the modified Bessel function of the second kind.

The parameter $\nu$ controls the mean-square differentiability of the RF: for any integer $k>0$, the field is $k$ times mean-square differentiable if and only if $\nu>k$.
The Mat\'ern model belongs to $\Phi_{\infty}$ and is therefore positive semidefinite in any Euclidean dimension.

The isotropic radial spectral density associated with~\eqref{eq:matern} in $\mathbb{R}^d$, normalized so that it integrates to one on $\mathbb{R}^d$, is
\begin{equation}\label{eq:matern-spectral}
g_{R}^{\mathcal{M}}(r)
= \frac{\Gamma(\nu+\tfrac{d}{2})\,\alpha^{d}}{\Gamma(\nu)\,\pi^{d/2}}\,
\bigl(1+\alpha^{2} r^{2}\bigr)^{-(\nu+d/2)},
\qquad r \ge 0,
\end{equation}
so that the full spectral density of the correlation function is
$
f(\boldsymbol{\omega}) = g_{R}^{\mathcal{M}}(\|\boldsymbol{\omega}\|).
$

A convenient feature of the Mat\'ern model is that its spectral distribution admits an exact Gaussian scale-mixture representation, which allows direct sampling of the spectral frequencies.
Let
\[
T \sim \mathrm{Gamma}(\nu,1),
\qquad
\mathbf{Z} \sim \mathcal{N}(\mathbf{0},I_d),
\qquad
T \perp\!\!\!\perp \mathbf{Z},
\]
and define
\begin{equation}\label{eq:matern-mixture}
\boldsymbol{\Omega} = \frac{1}{\alpha\sqrt{2T}}\,\mathbf{Z}
\;\in\mathbb{R}^d.
\end{equation}
Then $\boldsymbol{\Omega}$ follows a multivariate Student--$t$ distribution \citep{fang1990symmetric} with $2\nu$ degrees of freedom and scale matrix $(2\nu\alpha^{2})^{-1}I_d$, and has isotropic density
\[
f_{\boldsymbol{\Omega}}(\boldsymbol{\omega})
=
\frac{\Gamma(\nu+\tfrac{d}{2})\,\alpha^{d}}{\Gamma(\nu)\,\pi^{d/2}}
\bigl(1+\alpha^{2}\|\boldsymbol{\omega}\|^{2}\bigr)^{-(\nu+d/2)},
\]
which coincides with~\eqref{eq:matern-spectral}.
Hence, spectral frequencies for the Mat\'ern model can be sampled exactly by drawing $T$ and $\mathbf{Z}$ and setting $\boldsymbol{\Omega}$ as in~\eqref{eq:matern-mixture}.

The Mat\'ern correlation also admits the integral representation
\begin{equation}\label{eq:matern-integral}
\mathcal{M}_{\nu,\alpha}(x)
=
\int_{0}^{\infty}
\exp\!\Bigl(-\tfrac{x^{2}}{2\lambda}\Bigr)\,
p_{\nu,\alpha}(\lambda)\,{\rm d}\lambda,
\end{equation}
where $\lambda \sim \mathrm{Gamma}\!\bigl(\nu,2\alpha^{2}\bigr)$ has density
\[
p_{\nu,\alpha}(\lambda)
= \frac{1}{\Gamma(\nu)\,(2\alpha^{2})^{\nu}}\,
\lambda^{\nu-1}\exp\!\Bigl(-\tfrac{\lambda}{2\alpha^{2}}\Bigr),
\qquad \lambda>0.
\]
The representations~\eqref{eq:matern-mixture} and~\eqref{eq:matern-integral} are equivalent through the change of variable $\lambda=2\alpha^{2}T$.

This yields the following STB sampler for the covariance model $\sigma^2\mathcal{M}_{\nu,\alpha}$ \citep{EmeryLantu2006}:
\begin{enumerate}
\item Set the parameters $(\nu,\alpha,\sigma^2)$ and the number of spectral components $L$.
\item For $\ell=1,\ldots,L$:
\begin{enumerate}
\item Sample $T_\ell \sim \mathrm{Gamma}(\nu,1)$;
\item Sample $\mathbf{Z}_\ell \sim \mathcal{N}(\mathbf{0},I_d)$;
\item Set $\boldsymbol{\Omega}_\ell = \dfrac{1}{\alpha}\,\dfrac{\mathbf{Z}_\ell}{\sqrt{2T_\ell}}$;
\item Sample $\Phi_\ell \sim \mathrm{Unif}(0,2\pi)$;
\item Sample $\varepsilon_\ell\sim\mathrm{Unif}(0,1)$;
\end{enumerate}
\item Evaluate $\widetilde Z_L(\mathbf{s})$ at any target location $\mathbf{s}\in D$ via~\eqref{eq:TB-def-prob}.
\end{enumerate}

This construction provides exact sampling of the spectral variables entering the finite-$L$ STB representation for the covariance model $\sigma^2\mathcal{M}_{\nu,\alpha}$.
The resulting RF is therefore the usual finite-$L$ STB approximation to the target Gaussian field.

\section{STB simulation for the Kummer--Tricomi correlation model}
\label{sec4}

\cite{ma2022beyond} introduced the $\mathcal{K}$ family, which generalizes the Mat\'ern class by adding a parameter that controls the tail decay of spatial correlations.
This additional flexibility allows a substantially more independent tuning of smoothness and tail behavior, thereby providing a continuous bridge between the Mat\'ern model and heavier-tailed alternatives.

For parameters $\nu>0$ (smoothness), $\mu>0$ (tail), and $\beta>0$ (scale), the $\mathcal{K}$ correlation function is defined as
\begin{equation}\label{eq:kummer}
\mathcal{K}_{\nu,\mu,\beta}(x)
= \frac{\Gamma(\nu+\mu)}{\Gamma(\nu)}\,
U\!\left(\mu,\ 1-\nu,\ \frac{x^{2}}{2\beta^{2}}\right),
\qquad x \ge 0,
\end{equation}
where $U(a,b,z)$ denotes Tricomi's confluent hypergeometric function,
\[
U(a,b,z)
= \frac{1}{\Gamma(a)} \int_{0}^{\infty}
e^{-z t}\,t^{a-1}(1+t)^{b-a-1}\,{\rm d}t.
\]

For $\mu > d/2$, the model has an absolutely continuous isotropic spectral measure in $\mathbb{R}^d$, with radial spectral density given by \citep{Bhadra2025}
\begin{equation}\label{eq:kummer-spectral}
g_{R}^{\mathcal{K}}(r)
=
\frac{\Gamma\!\bigl(\nu+\tfrac d2\bigr)}{(2\pi)^{d/2}\,B(\mu,\nu)}\,
\beta^{d}\,
U\!\left(\nu+\tfrac d2,\ 1-\mu+\tfrac d2,\ \frac{\beta^{2}r^{2}}{2}\right),
\qquad r \ge 0,
\end{equation}
where $B(\mu,\nu)$ is the Beta function.
When $\mu \le d/2$, the correlation exhibits polynomial decay and long-range dependence, so that \eqref{eq:kummer-spectral} should no longer be interpreted as a Lebesgue density.
In that regime, the relevant object is the underlying spectral measure.

The function $\mathcal{K}_{\nu,\mu,\beta}$ belongs to $\Phi_\infty$ and is therefore positive semidefinite in any Euclidean dimension.
Under a suitable scaling, the Mat\'ern model is recovered as a limiting case when $\mu \to \infty$; specifically,
\begin{equation}\label{kkmm}
\mathcal{K}_{\nu,\mu,\beta\sqrt{2(\mu+1)}}\!\left(x\right)
\;\xrightarrow[\mu \to \infty]{}\;
\mathcal{M}_{\nu,\beta}(x)
\end{equation}
uniformly in $x$.

As shown by \cite{ma2022beyond}, $\mathcal{K}_{\nu,\mu,\beta}$ can be expressed as a continuous mixture of Mat\'ern kernels with random scale parameter:
\begin{equation}\label{eq:kummer-mixture}
\mathcal{K}_{\nu,\mu,\beta}(x)
= \mathbb{E}_{\Phi}\!\left[\mathcal{M}_{\nu,\Phi}(x)\right],
\qquad
\Phi^{2} \sim \mathrm{Inv\text{-}Gamma}\!\left(\mu,\ \tfrac{\beta^{2}}{2}\right),
\end{equation}
where $\mathcal{M}_{\nu,\Phi}$ denotes the Mat\'ern correlation~\eqref{eq:matern} with smoothness $\nu$ and random scale parameter $\Phi$.

The model also admits a convenient Gaussian scale-mixture representation.
Let
$
X \sim \Gamma(\nu,1)
$
and
$
Y \sim \Gamma(\mu,1),
$
with $X$ and $Y$ independent, and define
\[
T := \frac{X}{Y} \sim \mathrm{BetaPrime}(\nu,\mu).
\]
For $\mathbf{Z}\sim \mathcal{N}(\mathbf{0},I_d)$ independent of $(X,Y)$, set
\[
\boldsymbol{\Omega}
= \frac{1}{\beta}\,\frac{\mathbf{Z}}{\sqrt{T}}.
\]
For $\mu>d/2$, the random vector $\boldsymbol{\Omega}$ has isotropic density
\[
f_{\boldsymbol{\Omega}}(\boldsymbol{\omega})
=
\frac{\Gamma\!\bigl(\nu+\tfrac{d}{2}\bigr)}{(2\pi)^{d/2}B(\mu,\nu)}
\,\beta^{d}\,
U\!\left(\nu+\tfrac{d}{2},\,1-\mu+\tfrac{d}{2},\,\frac{\beta^{2}\|\boldsymbol{\omega}\|^{2}}{2}\right),
\]
which coincides with~\eqref{eq:kummer-spectral}.
More generally, the same Gaussian scale-mixture construction is well defined for all $\mu>0$: it yields the spectral density when $\mu>d/2$, and otherwise is understood in terms of the underlying spectral measure through
\begin{equation}\label{eq:kummer-cosine}
\mathbb{E}\!\left[\cos(\boldsymbol{\Omega}\cdot\mathbf{h})\right]
=
\mathcal{K}_{\nu,\mu,\beta}(\|\mathbf{h}\|),
\qquad \mathbf{h}\in\mathbb{R}^d.
\end{equation}
Thus, for all $\mu>0$, this construction provides the spectral random element entering the finite-$L$ STB representation.

The inverse-Gamma mixture in~\eqref{eq:kummer-mixture} and the Gaussian scale-mixture representation above are equivalent.
Indeed,
\[
\Phi = \frac{\beta}{\sqrt{Y}},
\qquad
T = \frac{X}{Y},
\]
so that the model can be interpreted either as a scale mixture of Mat\'ern correlations or as a Beta-prime mixture of Gaussian components.
This dual interpretation is useful for simulation, since it leads to a direct STB construction.

This yields the following STB sampler for the covariance model $\sigma^2\mathcal{K}_{\nu,\mu,\beta}$:
\begin{enumerate}
\item Set the parameters $(\nu,\mu,\beta,\sigma^2)$ and the number of spectral components $L$.
\item For $\ell=1,\dots,L$:
\begin{enumerate}
\item Sample $X_\ell \sim \Gamma(\nu,1)$ and $Y_\ell \sim \Gamma(\mu,1)$ independently, and set $T_\ell = X_\ell / Y_\ell$;
\item Sample $\mathbf{Z}_\ell \sim \mathcal{N}(\mathbf{0},I_d)$;
\item Set $\boldsymbol{\Omega}_\ell = \beta^{-1}\mathbf{Z}_\ell / \sqrt{T_\ell}$;
\item Sample $\Phi_\ell \sim \mathrm{Unif}(0,2\pi)$;
\item Sample $\varepsilon_\ell\sim\mathrm{Unif}(0,1)$;
\end{enumerate}
\item Evaluate $\widetilde Z_L(\mathbf{s})$ at any target location $\mathbf{s}\in D$ via~\eqref{eq:TB-def-prob}.
\end{enumerate}

This construction provides exact sampling of the spectral variables entering the finite-$L$ STB representation for the covariance model $\sigma^2\mathcal{K}_{\nu,\mu,\beta}$.
For $\mu\le d/2$, exactness is understood with respect to the underlying spectral measure rather than a Lebesgue spectral density.
The resulting RF is therefore the usual finite-$L$ STB approximation to the target Gaussian field.

\section{STB simulation for the Gauss--Hypergeometric correlation model}
\label{sec5}

The $\mathcal{GH}$ class \citep{emery2022gauss,hyp2025} represents one of the most general families of compactly supported correlation models currently available in spatial statistics.
It provides a unified analytical framework encompassing, as special cases, the generalized Wendland models \citep{gnei02,BFFP}. The main appeal of this construction lies in its ability to generate covariance functions that are both \emph{compactly supported}---leading to sparse covariance matrices---and \emph{highly flexible} in terms of smoothness and local behavior.

For parameters $\delta>\tfrac{d}{2}$, $\beta>\tfrac{d}{2}$, $\gamma>\tfrac{d}{2}$, and compact-support radius $a>0$, the $\mathcal{GH}$ correlation model is defined as \citep{emery2022gauss}
\begin{equation}\label{eq:GH-def}
\scalebox{0.9}{$
\mathcal{GH}_{\delta,\beta,\gamma,a}(x)
=
\left\{
\begin{array}{ll}
\displaystyle
K(\beta,\gamma,\delta,d)\left(1-\frac{x^2}{a^2}\right)^{\beta-\delta+\gamma-\tfrac d2-1}
{}_{2}F_{1}\!\left(\beta-\delta,\,\gamma-\delta;\,
\beta-\delta+\gamma-\tfrac d2;\,
1-\frac{x^2}{a^2}\right),
&  x \leq a, \\[10pt]
0, & x>a,
\end{array}
\right.
$}
\end{equation}
with
\[
K(\beta,\gamma,\delta,d)=
\frac{\Gamma(\beta-\tfrac d2)\,\Gamma(\gamma-\tfrac d2)}
{\Gamma(\beta-\delta+\gamma-\tfrac d2)\,\Gamma(\delta-\tfrac d2)}.
\]
Sufficient conditions for the validity of the model are
\begin{equation}\label{eq:GH-cond}
\delta > \frac{d}{2},
\qquad
2(\beta-\delta)(\gamma-\delta) \geq \delta,
\qquad
2(\beta+\gamma) \geq 6\delta + 1.
\end{equation}

Here ${}_{2}F_{1}$ denotes Gauss's hypergeometric function, which governs the polynomial structure of the model.
The isotropic \emph{radial spectral density} associated with~\eqref{eq:GH-def} is given by
\begin{equation}\label{eq:GH-spectral}
g_{R}^{\mathcal{GH}}(r)
=
L(\delta,\beta,\gamma)
\,a^{d}\;
{}_{1}F_{2}\!\left(\delta;\,\beta,\,\gamma;\,-\frac{(ar)^2}{4}\right),
\qquad r\ge 0,
\end{equation}
where ${}_{1}F_{2}$ is a generalized hypergeometric function and the normalizing constant is
\[
L(\delta,\beta,\gamma)=
\frac{
\Gamma(\delta)\,\Gamma(\beta-\tfrac d2)\,\Gamma(\gamma-\tfrac d2)
}{
2^{d}\pi^{d/2}\,
\Gamma(\delta-\tfrac d2)\,
\Gamma(\beta)\,
\Gamma(\gamma)
}.
\]

Following \cite{hyp2025}, we adopt the parameterization
\begin{equation}\label{eq:GH-reparam}
\delta = \tfrac{d+1}{2} + \nu, \qquad
\beta = \delta + \tfrac{\mu}{2}, \qquad
\gamma = \delta + \tfrac{\mu}{2} + l,
\end{equation}
and write
\begin{equation}\label{polo8}
\widetilde{\mathcal{GH}}_{\nu,\mu,l,a}(x) :=
\mathcal{GH}_{\delta,\;\delta+\mu/2,\;\delta+\mu/2+l,\;a}(x).
\end{equation}
Here $\nu > -\tfrac{1}{2}$ controls the differentiability at the origin,
$\mu>0$ determines the decay rate within the support,
$l \ge 0$ allows additional flexibility in the tail shape,
and $a>0$ retains its geometric meaning as the compact-support radius.

\cite{hyp2025} establishes the following sharp validity conditions for the parametrization~\eqref{polo8}.
Let $d$ be fixed, $\nu > -\frac{1}{2}$, $\delta = \frac{d+1}{2}+\nu$, $a > 0$, and $l \ge 0$. Then:
\begin{enumerate}
\item If $0 \le l \le \frac{d}{2}+\nu$, then
$
\widetilde{\mathcal{GH}}_{\nu,\mu,l,a} \in \Phi_d
\quad\text{if and only if}\quad
\mu \ge \delta - l + \frac{1}{2}.
$

\item If $l > \frac{d}{2}+\nu$, then
$
\widetilde{\mathcal{GH}}_{\nu,\mu,l,a} \in \Phi_d
\quad\text{if}\quad
\mu \ge \sqrt{2\nu + l^2 + d + 1} - l,
$
and only if
$
\mu \ge \delta - l + \frac{1}{2}.
$
\end{enumerate}

Two particularly relevant subfamilies arise from specific choices of $l$:
\begin{itemize}
\item \textbf{Hypergeometric model:}
$\mathcal{H}_{\nu,\mu,a} \equiv \widetilde{\mathcal{GH}}_{\nu,\mu,d/2+\nu,a}$, obtained for $l=\tfrac{d}{2}+\nu$.
This family, proposed in \cite{hyp2025}, has appealing theoretical properties; in particular, it allows one to maximize the \emph{integral range}, with the maximum attained at $\mu=1$.
It generalizes several well-known compactly supported correlation models, such as Euclid's hat, which in turn encompasses the classical triangular, circular, and spherical models \citep{Matheron1965, Gneiting1999}.

\item \textbf{Generalized Wendland model:}
$\mathcal{GW}_{\nu,\mu,a} \equiv \widetilde{\mathcal{GH}}_{\nu,\mu,1/2,a}$, obtained for $l=\tfrac{1}{2}$.
This celebrated class was first introduced by \cite{gnei02} and has been extensively studied in the spatial statistics and numerical analysis literature
\citep{Wen,Schaback:2011,Hubb14,BFFP,bevele}.
\end{itemize}

Both the $\mathcal{H}$ and $\mathcal{GW}$ models, after proper rescaling of the compact support,
allow one to recover the $\mathcal{M}$ model as a limiting case. For instance, \cite{bevilacqua2022unifying} show that
\begin{equation}\label{gwmm}
\mathcal{GW}_{\nu,\mu,\,
\beta\!\left(\frac{\Gamma(\mu+2\nu+1)}{\Gamma(\mu)}\right)^{\!\frac{1}{1+2\nu}}}(x)
\;\xrightarrow[\mu \to \infty]{}\;
\mathcal{M}_{\nu+\frac{1}{2},\,\beta}(x),
\end{equation}
uniformly in $x$.

\subsection{First scale mixture representation and simulation algorithm}

Under the conditions
\begin{equation}
\label{authorizedregion}
\mu>1, \qquad \frac{\mu}{2}-\frac{d}{2}-\frac{1}{2}-\nu+l>0,
\end{equation}
the $\widetilde{\mathcal{GH}}$ model admits a two-stage Beta mixture representation. By Theorem~6 in \cite{hyp2025},
\begin{equation}\label{eq:GH-mixture}
\widetilde{\mathcal{GH}}_{\nu,\mu,l,a}(x)
=
M(\nu,\mu,l,d)
\!\!\iint_{(0,1)^2}\!\!
u^{\delta-\frac{d+1}{2}}(1-u)^{\frac{\mu}{2}-\frac{3}{2}}
v^{\delta-\frac12+\nu}(1-v)^{\frac{\mu}{2}-\frac{d}{2}-\frac{3}{2}-\nu+l}
\mathcal{H}_{\nu,1,a\sqrt{uv}}(x)\,{\rm d}u\,{\rm d}v,
\end{equation}
where
\begin{equation}\label{eq:GH-constant}
M(\nu,\mu,l,d)
=
\frac{
\Gamma\!\bigl(\nu+\tfrac{\mu}{2}+\tfrac12\bigr)
\Gamma\!\bigl(\nu+\tfrac{\mu}{2}+\tfrac12+l\bigr)
}{
\Gamma(\nu+1)
\Gamma\!\bigl(\tfrac d2+2\nu+1\bigr)
\Gamma\!\bigl(\tfrac{\mu}{2}-\tfrac12\bigr)
\Gamma\!\bigl(\tfrac{\mu}{2}-\tfrac d2-\tfrac12-\nu+l\bigr)
}.
\end{equation}

Equivalently, if
\[
U \sim \mathrm{Beta}\!\left(1+\nu,\tfrac{\mu}{2}-\tfrac12\right),
\qquad
V \sim \mathrm{Beta}\!\left(\tfrac d2+2\nu+1,\tfrac{\mu}{2}-\tfrac d2-\tfrac12-\nu+l\right),
\]
independently, then
\[
\widetilde{\mathcal{GH}}_{\nu,\mu,l,a}(x)
=
\mathbb{E}_{U,V}\!\left[\mathcal{H}_{\nu,1,a\sqrt{UV}}(x)\right].
\]
Hence, the model is represented as a scale mixture of compactly supported base kernels.

The corresponding conditional radial spectral density of the base kernel $\mathcal{H}_{\nu,1,b}$ is
\begin{equation}\label{eq:base-spectral}
g^{\widetilde{\mathcal{GH}}}_{R}(r\mid b)
=
C_H(\nu,d)\,b^d\,
{}_1F_2\!\left(\delta;\delta+\tfrac12,2\delta;-\frac{(br)^2}{4}\right),
\qquad r\ge 0,
\end{equation}
where
\begin{equation}\label{eq:CH-corrected}
C_H(\nu,d)=
\frac{
\Gamma\!\bigl(\tfrac{d+1}{2}+\nu\bigr)
\Gamma(1+\nu)
\Gamma\!\bigl(\tfrac d2+1+2\nu\bigr)
}{
2^d\pi^{d/2}
\Gamma\!\bigl(\tfrac12+\nu\bigr)
\Gamma\!\bigl(\tfrac d2+1+\nu\bigr)
\Gamma(d+1+2\nu)
}.
\end{equation}
Using Watson’s identity \citep[p.~147]{Watson}, this can be rewritten as
\begin{equation}\label{eq:base-bessel}
g^{\widetilde{\mathcal{GH}}}_{R}(r\mid b)
=
C_H(\nu,d)\,b^d\,
\Gamma^2\!\left(\delta+\tfrac12\right)
\left(\frac{br}{4}\right)^{1-2\delta}
J_{\delta-\tfrac12}^2\!\left(\frac{br}{2}\right).
\end{equation}

Let $T=bR$. Then the distribution of $T$ does not depend on $b$, and its density is
\begin{equation}\label{eq:fT-bessel}
\begin{aligned}
f_T(t)
&=
|\mathbb{S}^{d-1}|\,t^{d-1}\,C_H(\nu,d)\,
{}_1F_2\!\left(\delta;\delta+\tfrac12,2\delta;-\tfrac{t^2}{4}\right)\\
&=
|\mathbb{S}^{d-1}|\,t^{d-1}\,C_H(\nu,d)\,
\Gamma^2\!\left(\delta+\tfrac12\right)
\left(\tfrac{t}{4}\right)^{1-2\delta}
J_{\delta-\tfrac12}^2\!\left(\tfrac{t}{2}\right),
\qquad t\ge0.
\end{aligned}
\end{equation}
This yields a convenient factorization of the spectral radius:
first sample $T$ from the universal density~\eqref{eq:fT-bessel}, then set
\[
R=\frac{T}{a\sqrt{UV}}.
\]

This leads to the following STB simulation scheme for the covariance model
$\sigma^2\widetilde{\mathcal{GH}}_{\nu,\mu,l,a}$.

\begin{enumerate}
\item Fix $(\nu,\mu,l,a,\sigma^2)$ and the number of spectral components $L$.
\item For $\ell=1,\ldots,L$:
\begin{enumerate}
\item sample
\[
U_\ell \sim \mathrm{Beta}\!\left(1+\nu,\tfrac{\mu}{2}-\tfrac12\right),
\qquad
V_\ell \sim \mathrm{Beta}\!\left(\tfrac d2+2\nu+1,\tfrac{\mu}{2}-\tfrac d2-\tfrac12-\nu+l\right),
\]
independently;
\item set $b_\ell=a\sqrt{U_\ell V_\ell}$;
\item sample $T_\ell$ from~\eqref{eq:fT-bessel};
\item set $R_\ell=T_\ell/b_\ell$;
\item sample $\boldsymbol{\Theta}_\ell\sim\mathrm{Unif}(\mathbb{S}^{d-1})$ and define $\boldsymbol{\Omega}_\ell=R_\ell\boldsymbol{\Theta}_\ell$;
\item sample $\Phi_\ell\sim\mathrm{Unif}(0,2\pi)$ and $\varepsilon_\ell\sim\mathrm{Unif}(0,1)$.
\end{enumerate}
\item Evaluate $\widetilde Z_L(\mathbf{s})$ using~\eqref{eq:TB-def-prob}.
\end{enumerate}

The main numerical task is the generation of the auxiliary variable $T$.
Since the density~\eqref{eq:fT-bessel} is oscillatory and does not admit a closed-form inverse cdf, we use a two-piece acceptance--rejection sampler on $(0,\infty)$.
Given a threshold $t_0>0$, the proposal density is
\[
g(t)=w_1 g_1(t)+w_2 g_2(t),
\]
where
\begin{align}
g_1(t) &= \frac{d\,t^{d-1}}{t_0^d}, && 0\le t\le t_0,\\
g_2(t) &= \frac{\kappa\,t_0^\kappa}{t^{\kappa+1}}, && t>t_0,
\end{align}
with
$
\kappa=2\nu+1.
$
The first component matches the behavior of $f_T$ near the origin, whereas the second reproduces its polynomial tail decay. Sampling from both proposals is explicit:
\[
T=t_0U^{1/d}\quad\text{under }g_1,
\qquad
T=t_0(1-U)^{-1/\kappa}\quad\text{under }g_2,
\qquad U\sim\mathrm{Unif}(0,1).
\]

For numerical implementation, the threshold $t_0$ is selected by minimizing the sum of the body and tail envelope constants over a bounded interval.
A small multiplicative safety factor is incorporated into the envelopes to guarantee domination under floating-point arithmetic.
This yields a stable rejection sampler that does not require numerical inversion of the target cdf.

The Bessel term in~\eqref{eq:fT-bessel} is evaluated using a piecewise strategy:
a series expansion for small arguments, the standard library routine in the intermediate regime, and an asymptotic approximation for large arguments.
This avoids loss of accuracy both near the origin and in the tail and improves the robustness of the rejection sampler.

Under~\eqref{authorizedregion}, the Beta variables are non-degenerate and the mixture sampler applies directly.
In the implementation, the boundary cases are also handled by continuous extension:
when $\mu$ is numerically close to $1$, $U$ is set to $1$, and when the second shape parameter of $V$ is numerically close to zero, $V$ is set to $1$.
Thus, part of the boundary of the simulatability region can be treated in practice, although the integral representation~\eqref{eq:GH-mixture} is stated under strict inequalities.

\begin{figure}[ht!]
    \centering
    \includegraphics[width=0.32\textwidth]{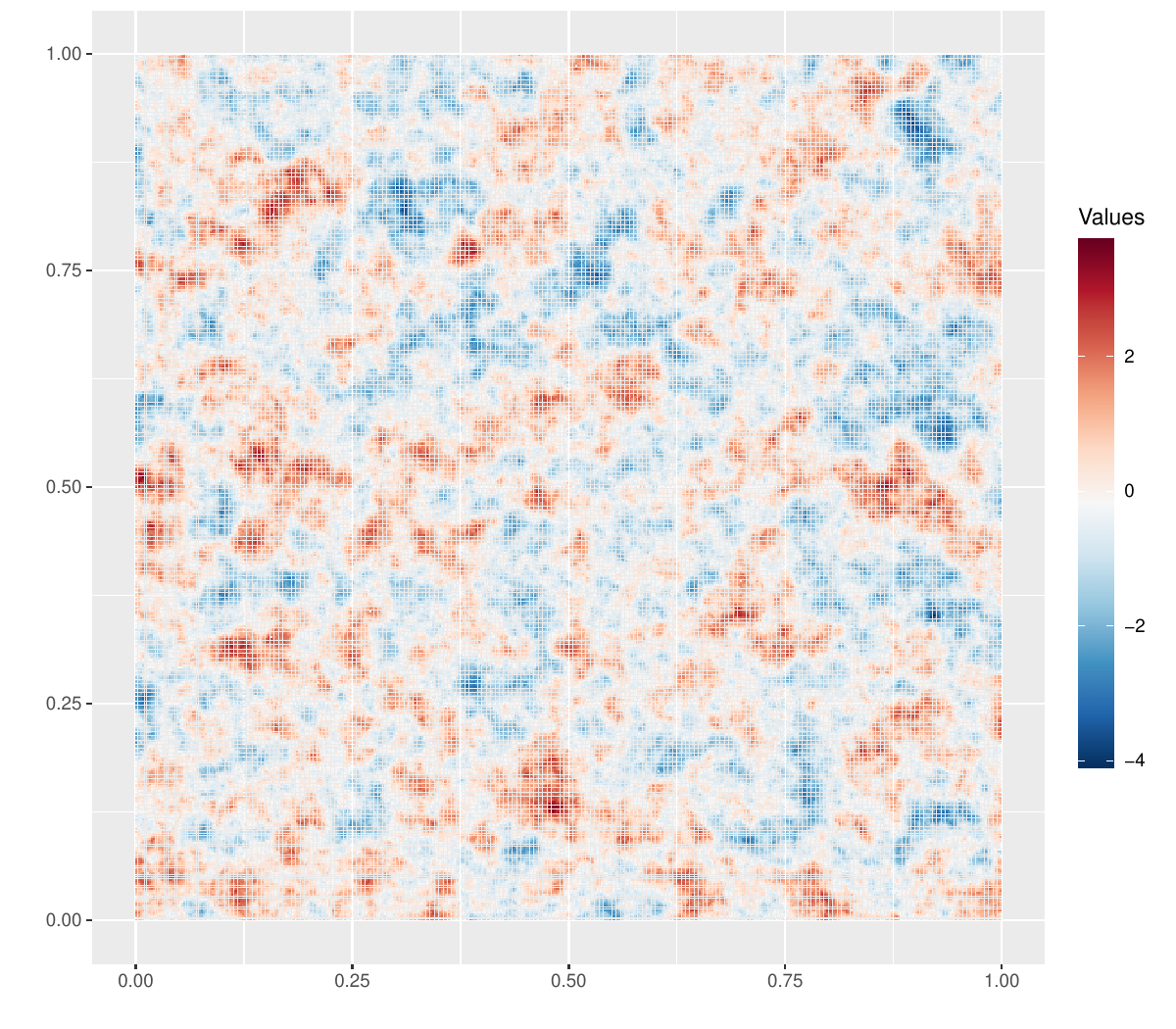}\hfill
    \includegraphics[width=0.32\textwidth]{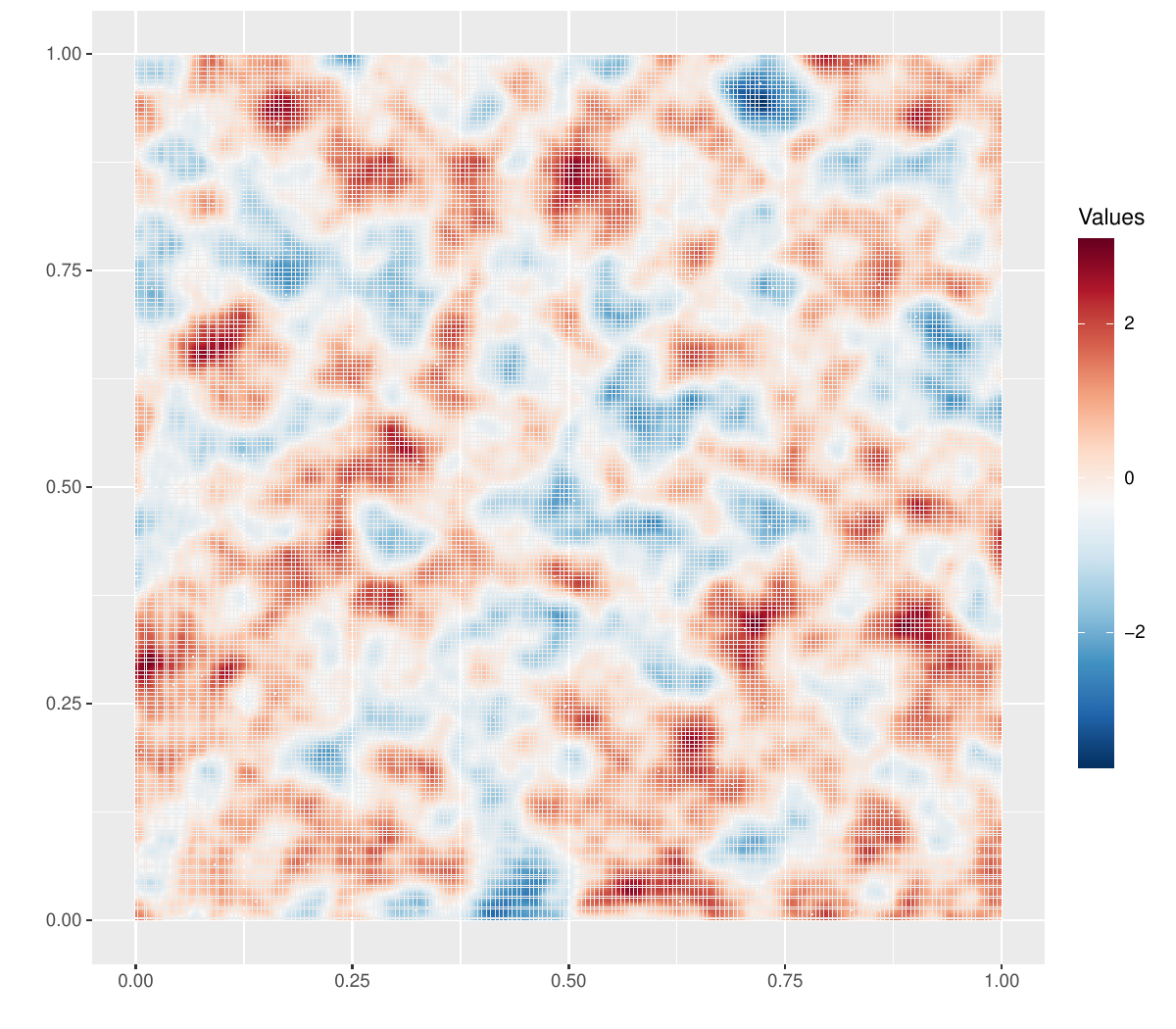}\hfill
    \includegraphics[width=0.32\textwidth]{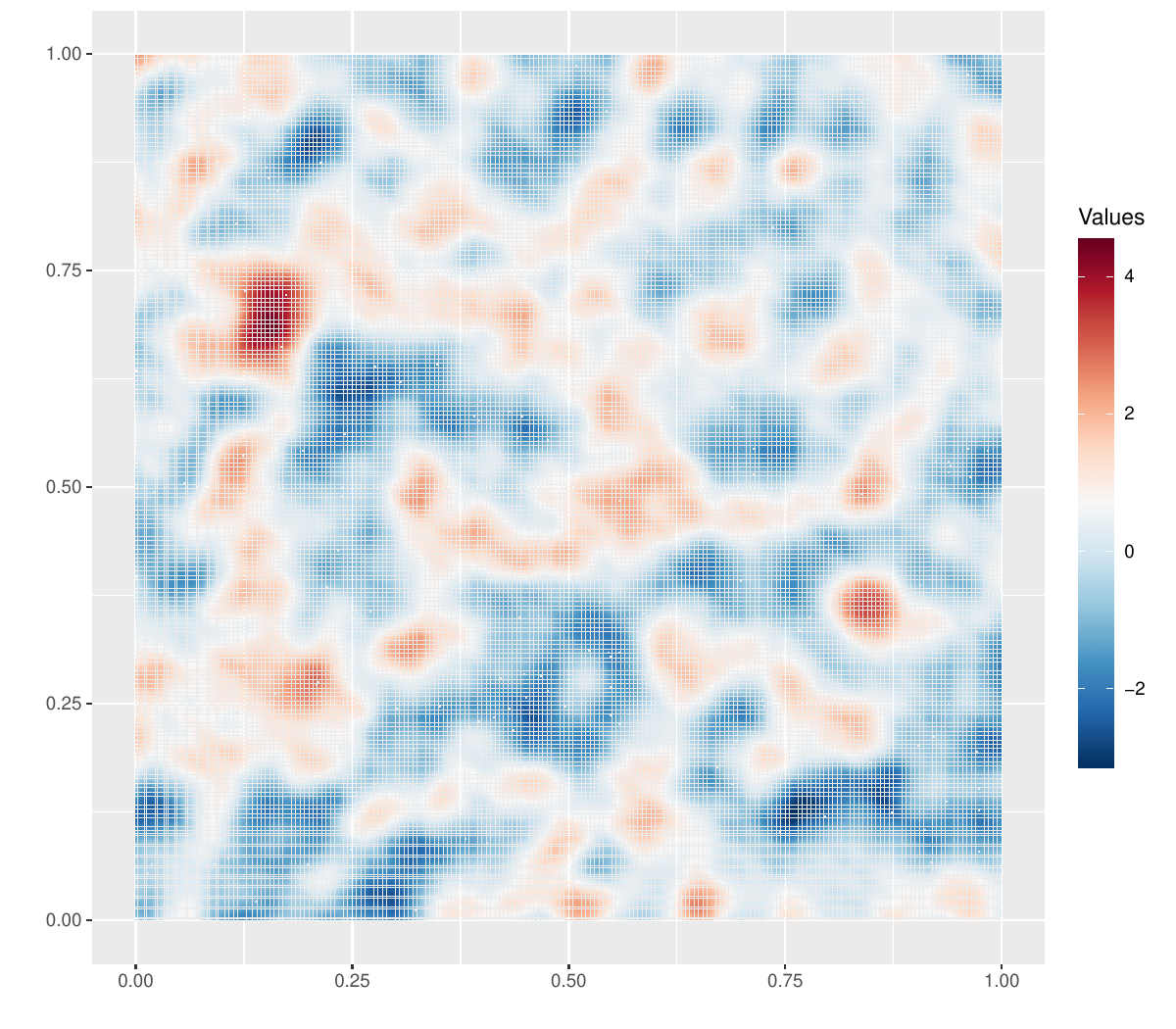}
    \caption{Simulated Gaussian RFs on $1{,}000{,}000$ points uniformly distributed on $[0,1]^2$ using the generalized Wendland model $\mathcal{GW}_{\nu,7,0.2}\equiv\widetilde{\mathcal{GH}}_{\nu,7,1/2,0.2}$, for $\nu=0,1,2$ from left to right.}
    \label{fig:g-simulations}
\end{figure}

Figure~\ref{fig:g-simulations} shows three realizations on $1{,}000{,}000$ points uniformly distributed on $[0,1]^2$, obtained with $L=1{,}000$ under the generalized Wendland model
$
\mathcal{GW}_{\nu,7,0.2}\equiv \widetilde{\mathcal{GH}}_{\nu,7,1/2,0.2},
$
for $\nu=0,1,2$.

The conditions in~\eqref{authorizedregion} define the parameter region where the Beta--mixture construction is directly available.
They are more restrictive than the validity conditions of the covariance model.
Therefore, every model covered by the sampler is valid, but not every valid model is covered by this first representation.

\begin{figure}[h!]
    \centering
    \includegraphics[width=0.45\textwidth]{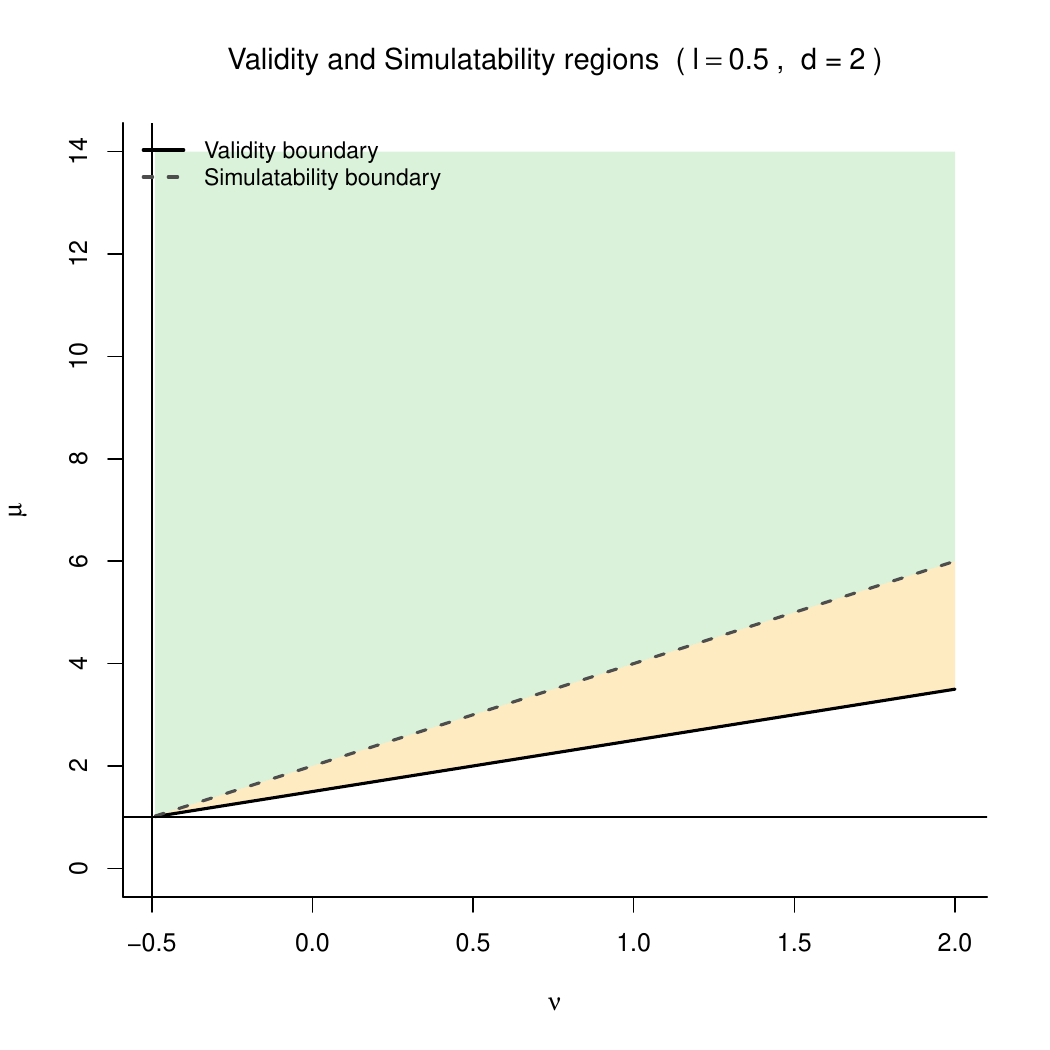}\hfill
    \includegraphics[width=0.45\textwidth]{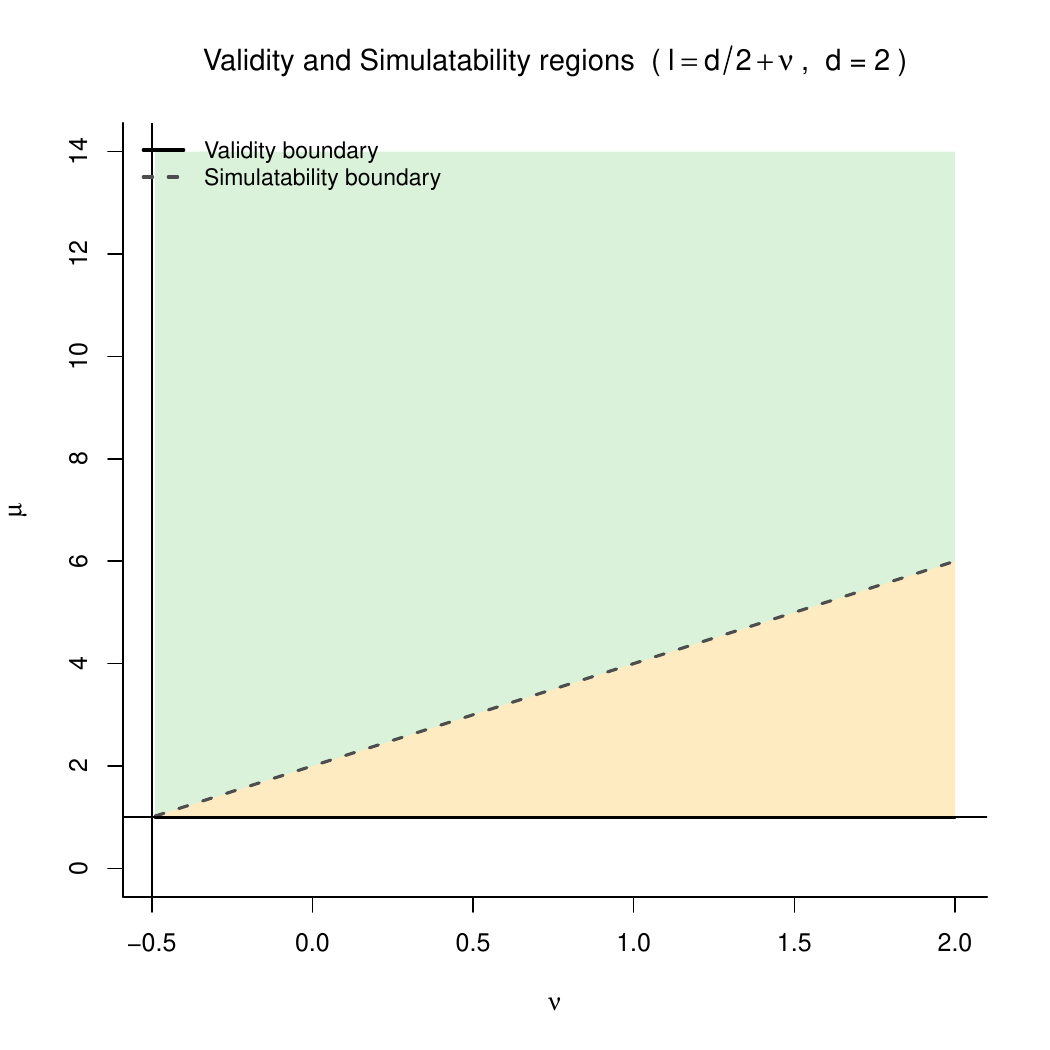}
    \caption{Validity (yellow and green) and simulatability (green) regions in the $(\nu,\mu)$ plane for $d=2$. The left panel corresponds to the $\mathcal{GW}$ model ($l=0.5$), while the right panel refers to the $\mathcal{H}$ model ($l=d/2+\nu$). The solid black line denotes the validity boundary, whereas the dashed black line marks the simulatability limit of the Beta--mixture construction.}
    \label{fig:validity-simulability}
\end{figure}

The gap between validity and direct simulatability is illustrated in Figure~\ref{fig:validity-simulability}, which reports the $(\mu,\nu)$ regions for $d=2$.
The left panel corresponds to the $\mathcal{GW}$ model ($l=0.5$), whereas the right panel corresponds to the $\mathcal{H}$ model ($l=d/2+\nu$).
In both cases, the simulatability region is strictly contained in the validity region.
The uncovered region is larger for the $\mathcal{H}$ model, indicating a narrower parameter range for which the Beta--mixture representation can be used directly.

As in the next subsection, exactness refers only to the sampling of the spectral variables entering the finite-$L$ STB representation.
The resulting RF is therefore the usual finite-$L$ STB approximation to the target Gaussian field.

\subsection{Second scale mixture representation and simulation algorithm}

To cover the parameter region not handled by the Beta--mixture construction, we use an alternative spectral decomposition based on Gasper's expansion of the hypergeometric kernel.
Specifically, using the original parameterization $(\delta,\beta,\gamma)$, the radial spectral density~\eqref{eq:GH-spectral} of the $\mathcal{GH}$ kernel can be rewritten as \citep[Eq.~3.1]{Gasper:75}
\begin{equation*}
\begin{split}
g_{R}^{\mathcal{GH}}(r)
&= L(\delta,\beta,\gamma)\,a^d\,\Gamma^2(\eta+1)
\left(\frac{ar}{4}\right)^{-2\eta}
\sum_{n=0}^{\infty}
C(n,\delta,\beta,\gamma)
\frac{(2n+2\eta)(2\eta+1)_n}{(n+2\eta)n!}
J^2_{\eta+n}\!\left(\frac{ar}{2}\right),
\end{split}
\end{equation*}
where $(\cdot)_n$ denotes the Pochhammer symbol,
$
\eta = \frac{1}{2}\left(\beta+\gamma-\delta-\frac{3}{2}\right),
$
and
\begin{equation*}
C(n,\delta,\beta,\gamma)
=
{}_4F_3\!\left(-n,n+2\eta,\eta+1,\delta;\eta+\tfrac12,\beta,\gamma;1\right).
\end{equation*}
Since the hypergeometric function ${}_4F_3$ is terminating, it can be evaluated recursively as
\[
t_{k+1}
=
t_k\,
\frac{(-n+k)(n+2\eta+k)(\eta+1+k)(\delta+k)}
{(\eta+\tfrac12+k)(\beta+k)(\gamma+k)(k+1)},
\qquad
t_0=1,
\qquad
C(n,\delta,\beta,\gamma)=\sum_{k=0}^n t_k.
\]

Define
\begin{equation}
\label{eq:holeeffectdensity}
\begin{split}
g^{(n)}_R\!\left(r \mid \eta+n,a\right)
&= C_H(\eta+n-d/2,d,n)\, a^{d+2n} r^{2n} \\
&\quad \times
{}_1F_2\!\left(
\tfrac{d+1}{2}+\eta+n-\tfrac d2;\,
\tfrac d2+1+\eta+n-\tfrac d2,\,
1+2\eta+2n;\,
-\tfrac{(ar)^2}{4}
\right).
\end{split}
\end{equation}
By Watson's identity, this can be rewritten as
\begin{equation}
\label{eq:holeeffectdensity-bessel}
g^{(n)}_R\!\left(r \mid \eta+n,a\right)
=
C_H(\eta+n-d/2,d,n)\, a^{d+2n} r^{2n}
\Gamma^2(1+\eta+n)
\left(\tfrac{ar}{4}\right)^{-2\eta-2n}
J^2_{\eta+n}\!\left(\tfrac{ar}{2}\right).
\end{equation}
Therefore,
\begin{equation}
\label{eq:gasper-mixture}
g_R^{\mathcal{GH}}(r)
=
\sum_{n=0}^{\infty}
w_n(\delta,\beta,\gamma)\,
g^{(n)}_R\!\left(r \mid \eta+n,a\right),
\end{equation}
where
\begin{equation}\label{polo}
w_n(\delta,\beta,\gamma)
=
\frac{L(\delta,\beta,\gamma)\Gamma^2(\eta+1)}{2^{4n}\Gamma^2(\eta+n+1)}
\frac{C(n,\delta,\beta,\gamma)}{C_H(\eta+n-\frac d2,d,n)}
\frac{(2n+2\eta)(2\eta+1)_n}{(n+2\eta)n!}.
\end{equation}

Under conditions~\eqref{eq:GH-cond}, \citet[Lemma~3.1]{cho2020rational} proved that
$C(n,\delta,\beta,\gamma)\ge 0$.
Since both $g_R^{\mathcal{GH}}$ and $g_R^{(n)}(\cdot\mid\eta+n,a)$ integrate to one on $\mathbb{R}^d$,
it follows that the weights $w_n(\delta,\beta,\gamma)$ are non-negative and sum to one.
Hence, \eqref{eq:gasper-mixture} is a genuine discrete mixture representation of the spectral density.
In particular, the Gasper--mixture formulation applies on the whole validity region of the $\widetilde{\mathcal{GH}}$ model and therefore complements the Beta--mixture construction.

For each fixed $n$, let $\boldsymbol{\Omega}\in\mathbb{R}^d$ have radial density
$g_R^{(n)}(\cdot\mid\eta+n,a)$, let $R=\|\boldsymbol{\Omega}\|$, and define $T:=aR$.
Then the distribution of $T$ does not depend on $a$, and its density is
\begin{equation}\label{eq:fT-holeeffect}
\begin{aligned}
f_T^{(n)}(t)
&=
|\mathbb{S}^{d-1}|\,
C_H(\eta+n-\tfrac d2,d,n)\,
4^{2\eta+2n}\,
\Gamma^2(1+\eta+n)\,
t^{d-1-2\eta}
J^2_{\eta+n}\!\left(\tfrac{t}{2}\right),
\qquad t>0.
\end{aligned}
\end{equation}
Because $J_{\eta+n}(t)$ behaves like $t^{\eta+n}$ near zero and like a bounded oscillatory term times $t^{-1/2}$ for large $t$, the density $f_T^{(n)}$ is non-negative and bounded on $(0,\infty)$.

In practice, the infinite mixture in~\eqref{eq:gasper-mixture} is approximated by truncation at a fixed level $n_{\max}=15$.
More precisely, we compute the weights $\{w_n(\delta,\beta,\gamma)\}_{n=0}^{n_{\max}}$, evaluate them in log-scale for numerical stability, and then renormalize them to obtain a discrete distribution on $\{0,\dots,n_{\max}\}$.
This avoids underflow in the computation of the coefficients and proved sufficient in all numerical experiments reported in this paper.
With this approximation, the discrete mixture step has negligible computational overhead relative to the generation of the auxiliary variable $T$.

Based on this representation, the following STB simulation scheme can be used for the covariance model $\sigma^2 \widetilde{\mathcal{GH}}_{\nu,\mu,l,a}$ whenever the Beta--mixture sampler is not applicable.
As in the previous subsection, exactness here refers to the sampling of the spectral variables entering the STB construction.

\begin{enumerate}
\item Set the parameters $(\nu,\mu,l,a,\sigma^2)$ and the number of spectral components $L$.
\item Set $\delta = \tfrac{d+1}{2} + \nu$, $\beta = \delta + \tfrac{\mu}{2}$, and $\gamma = \delta + \tfrac{\mu}{2} + l$.
\item Compute the truncated weights $\{w_n(\delta,\beta,\gamma)\}_{n=0}^{n_{\max}}$ using~\eqref{polo}, and renormalize them.
\item For $\ell=1,\ldots,L$:
\begin{enumerate}
\item Sample $N_\ell$ from the discrete distribution on $\{0,\dots,n_{\max}\}$ with probabilities proportional to $w_n(\delta,\beta,\gamma)$;
\item Sample $T_\ell$ from the density $f_T^{(N_\ell)}$ defined in~\eqref{eq:fT-holeeffect};
\item Set the spectral radius as $R_\ell=T_\ell/a$;
\item Sample $\boldsymbol{\Theta}_\ell\sim\mathrm{Unif}(\mathbb{S}^{d-1})$ and define $\boldsymbol{\Omega}_\ell = R_\ell\,\boldsymbol{\Theta}_\ell$;
\item Sample $\Phi_\ell\sim\mathrm{Unif}(0,2\pi)$;
\item Sample $\varepsilon_\ell\sim\mathrm{Unif}(0,1)$.
\end{enumerate}
\item Evaluate $\widetilde Z_L(\mathbf{s})$ at any target location $\mathbf{s}$ using~\eqref{eq:TB-def-prob}.
\end{enumerate}

As in the Beta--mixture case, the main numerical ingredient is the sampling of the auxiliary variable $T$.
The density $f_T^{(n)}$ is oscillatory and does not admit a closed-form inverse cdf, so we use a two-piece acceptance--rejection scheme on $(0,\infty)$.
The support is partitioned at a threshold $t_0$ into a body and a tail region, with proposal densities
\[
g_1(t)=\frac{d\,t^{d-1}}{t_0^d},
\qquad 0\le t\le t_0,
\]
and
\[
g_2(t)=\frac{\alpha\,t_0^\alpha}{t^{\alpha+1}},
\qquad t>t_0,
\]
where
$
\alpha = 2\eta+1-d.
$
Hence, the tail proposal reproduces the polynomial decay of $f_T^{(n)}$.
Sampling from these proposals is explicit:
\[
T=t_0U^{1/d}\quad\text{under }g_1,
\qquad
T=t_0(1-U)^{-1/\alpha}\quad\text{under }g_2,
\qquad U\sim\mathrm{Unif}(0,1).
\]

The threshold $t_0$ is selected numerically by minimizing the sum of the corresponding envelope constants over a bounded interval.
To ensure domination in finite precision arithmetic, a small safety factor is incorporated into the envelope construction.
The resulting proposal yields stable acceptance rates across the parameter settings considered in our experiments.

The recursive evaluation of the terminating ${}_4F_3$ series and the log-scale computation of the weights are important implementation details.
They make the discrete mixture step numerically stable even for moderately large values of $n$.
Likewise, the Bessel evaluations entering~\eqref{eq:fT-holeeffect} are stabilized by combining a small-argument expansion, the standard library routine for intermediate values, and a large-argument asymptotic approximation.

In our implementation, the overall workflow is automatic:
whenever the Beta--mixture conditions are satisfied numerically, the Beta--mixture sampler is used; otherwise, the algorithm switches to the Gasper--mixture sampler.
The Gasper construction therefore complements the Beta one and extends the simulation procedure to the part of the admissible parameter region that is not practically covered by the Beta--mixture representation.

\section{Numerical experiments}
\label{sec6}

Throughout this section, we consider a zero-mean Gaussian random field
$\{Z(\mathbf{s}) : \mathbf{s}\in D\subset\mathbb{R}^2\}$ with unit variance, that is,
$\mathbb{E}[Z(\mathbf{s})]=0$ and $\mathbb{V}[Z(\mathbf{s})]=\sigma^2=1$.
The aim is to assess the empirical performance of the proposed STB algorithms in terms of second-order accuracy across a range of covariance models and parameter settings.

Unless otherwise stated, each experiment is based on $1{,}000$ independent realizations generated at $5{,}000$ spatial locations drawn independently from the uniform distribution on $[0,1]^2$.
All STB simulations are carried out with $L=1{,}000$ spectral components.

We first consider the generalized Wendland model
$
\mathcal{GW}_{\nu,\mu,a}\equiv \widetilde{\mathcal{GH}}_{\nu,\mu,1/2,a}.
$
The parameter settings are chosen to represent a range of smoothness and support configurations.
Since $\nu$ controls the mean-square differentiability of the field, we consider $\nu=0$ for non-differentiable realizations and $\nu=1$ for smoother realizations.
We also consider two compact-support parameters, $a=0.1$ and $a=0.5$, which determine the correlation range.

The shape parameter $\mu$ determines which of the two simulation algorithms applies.
We take $\mu=6,7$ for the Beta--mixture algorithm (Section~5.1), and $\mu=2,3$ for the Gasper--mixture algorithm (Section~5.2).
This yields eight scenarios, summarized in Tables~\ref{tab:tabla1} and~\ref{tab:tabla2}.

\begin{figure}[htbp]
    \centering
    \begin{subfigure}{0.45\textwidth}
        \centering
        \includegraphics[width=\linewidth]{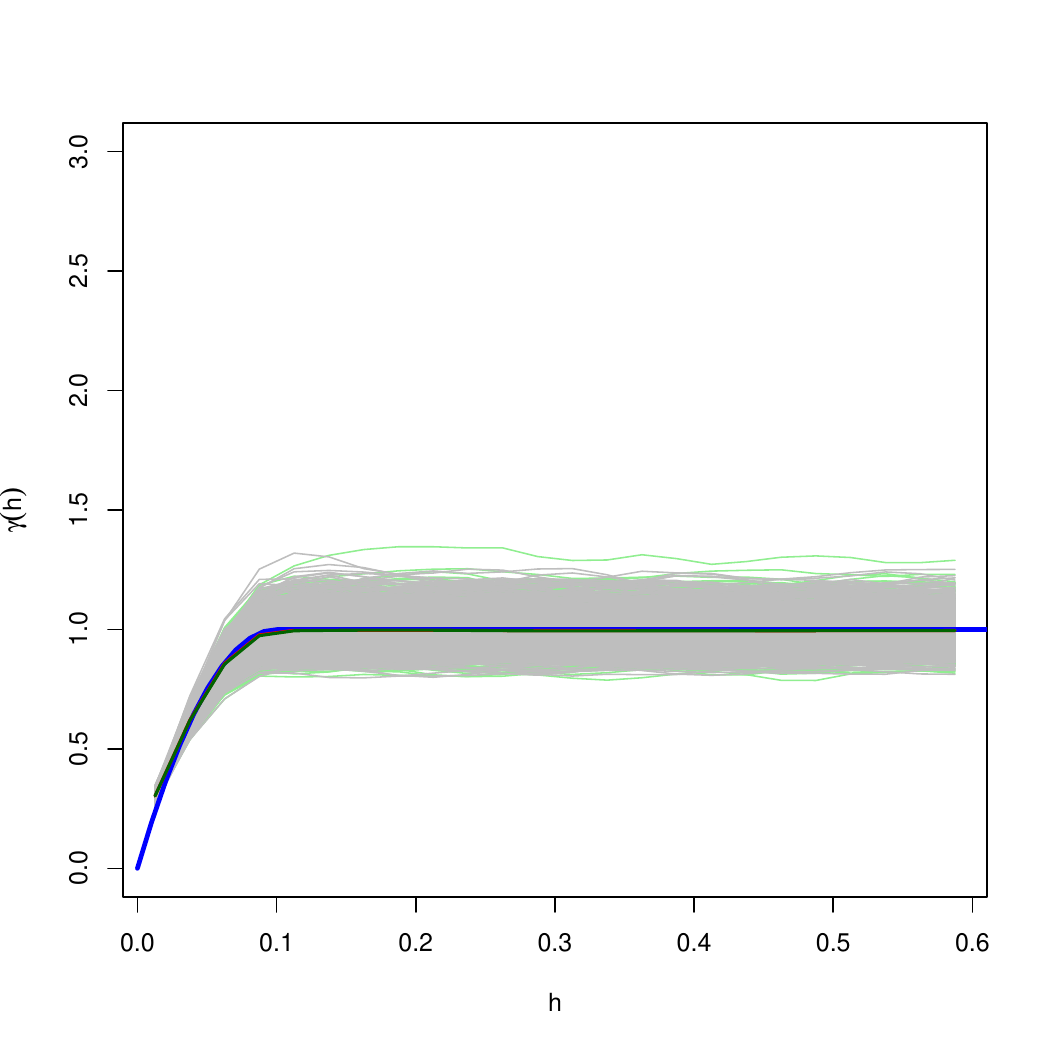}
        \caption{Scenario 1}
        \label{fig:1}
    \end{subfigure}
    \hfill
    \begin{subfigure}{0.45\textwidth}
        \centering
        \includegraphics[width=\linewidth]{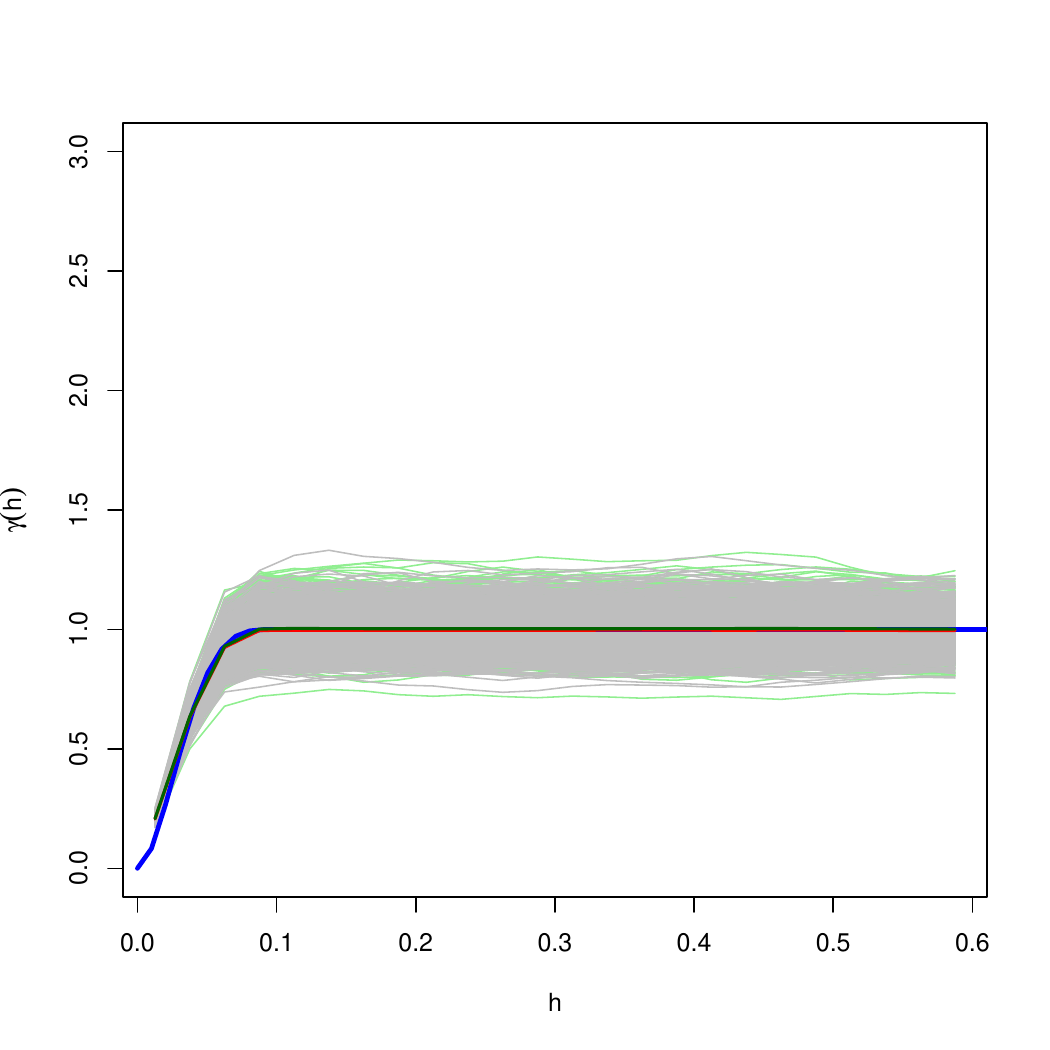}
        \caption{Scenario 2}
        \label{fig:2}
    \end{subfigure}
    \vspace{0.5cm}
    \begin{subfigure}{0.45\textwidth}
        \centering
        \includegraphics[width=\linewidth]{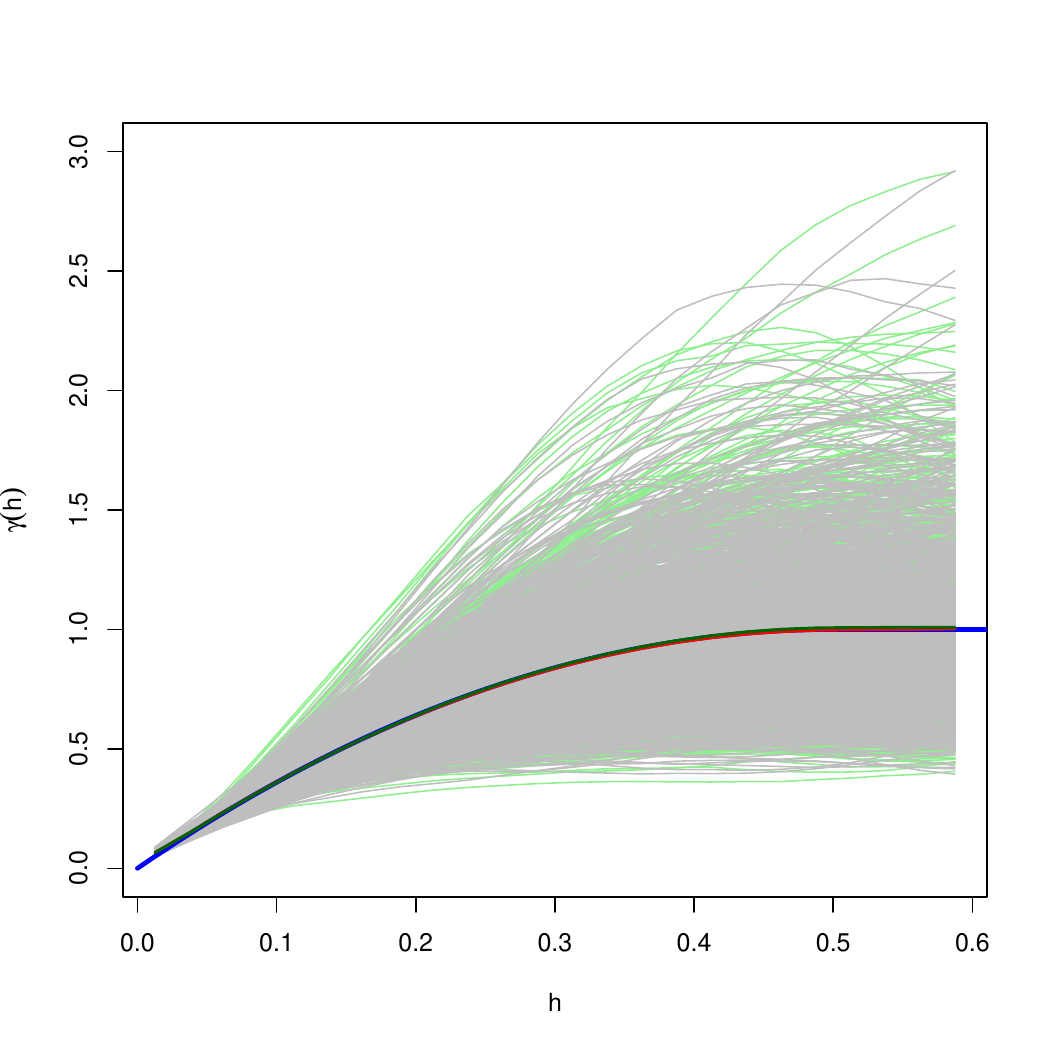}
        \caption{Scenario 3}
        \label{fig:3}
    \end{subfigure}
    \hfill
    \begin{subfigure}{0.45\textwidth}
        \centering
        \includegraphics[width=\linewidth]{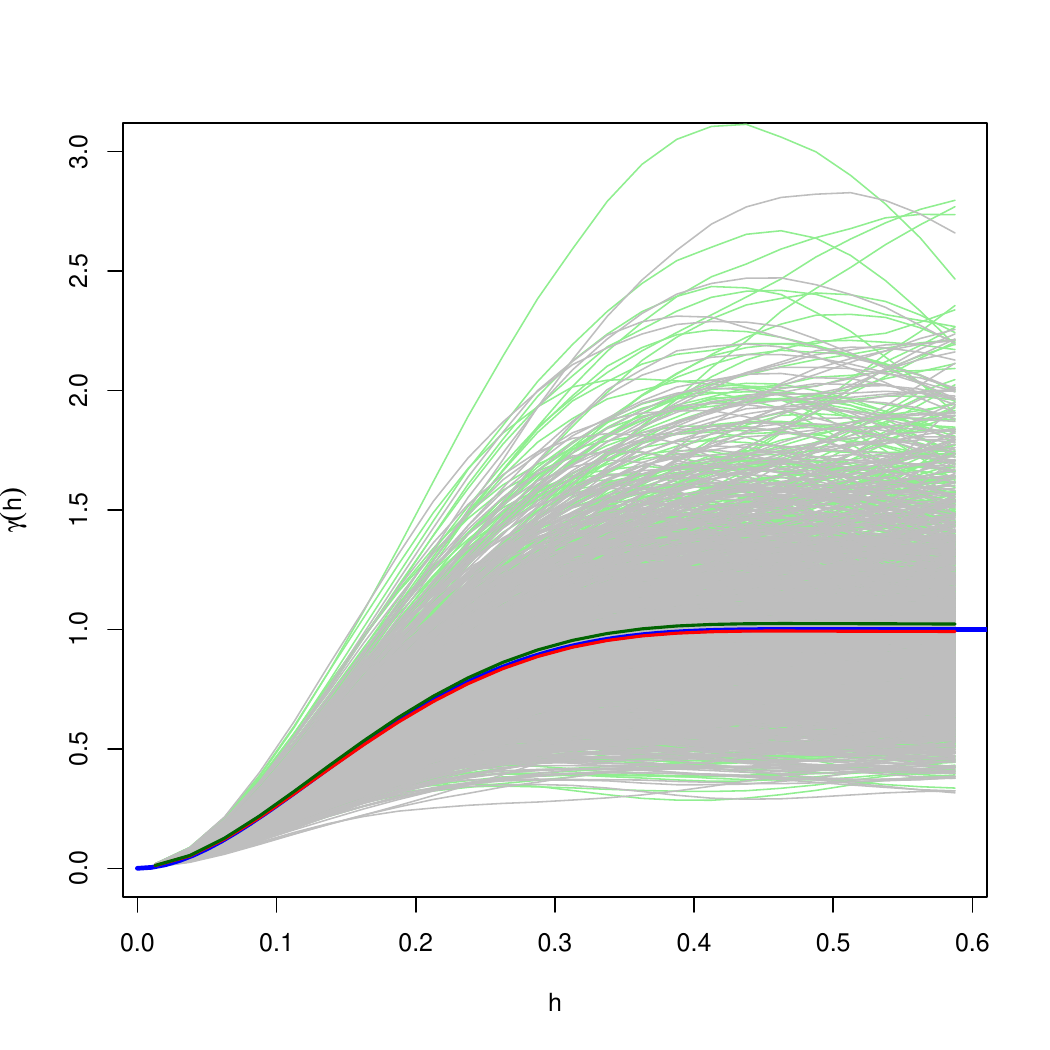}
        \caption{Scenario 4}
        \label{fig:4}
    \end{subfigure}

    \caption{Empirical semivariograms from $1{,}000$ realizations of a Gaussian random field under different parameter configurations of the $\mathcal{GW}_{\nu,\mu,a}$ model (see Table~\ref{tab:tabla1}), generated using the Beta--mixture algorithm. Gray lines show the empirical semivariograms for individual realizations, red lines their empirical mean over the $1{,}000$ realizations, and blue lines the corresponding theoretical semivariograms. Green lines show empirical semivariograms computed from realizations obtained via the Cholesky method.}
    \label{fig:simu1}
\end{figure}

\begin{figure}[htbp]
    \centering
    \begin{subfigure}{0.45\textwidth}
        \centering
        \includegraphics[width=\linewidth]{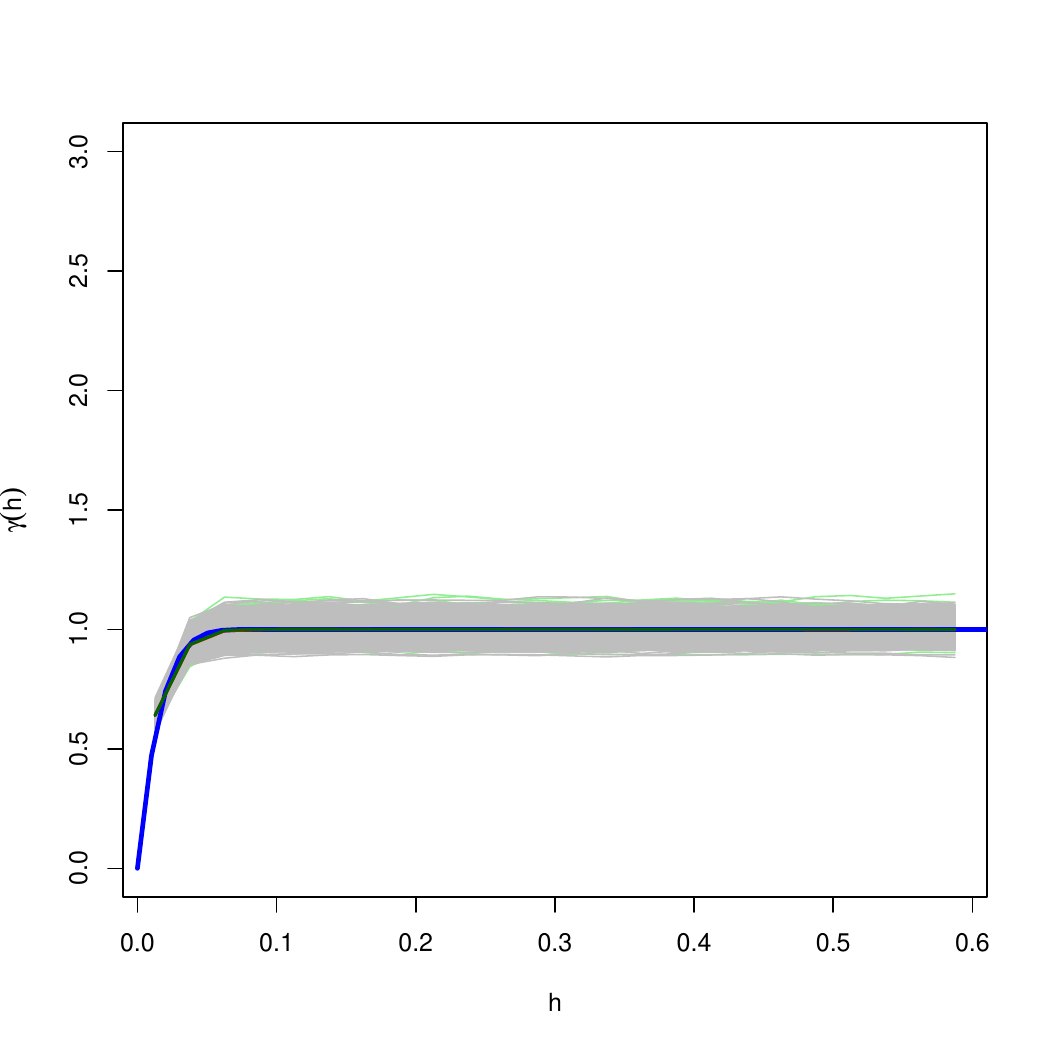}
        \caption{Scenario 5}
        \label{fig:5}
    \end{subfigure}
    \hfill
    \begin{subfigure}{0.45\textwidth}
        \centering
        \includegraphics[width=\linewidth]{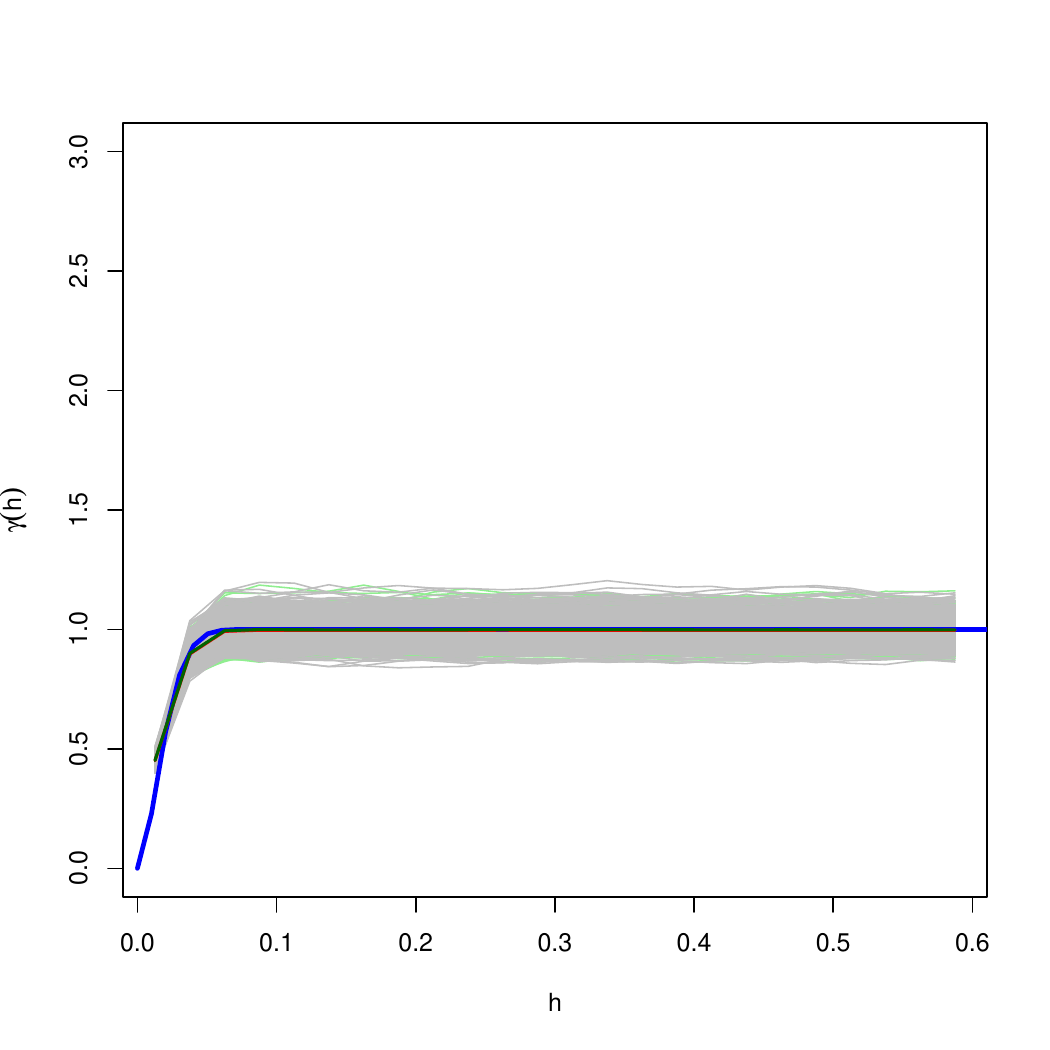}
        \caption{Scenario 6}
        \label{fig:6}
    \end{subfigure}
    \vspace{0.5cm}
    \begin{subfigure}{0.45\textwidth}
        \centering
        \includegraphics[width=\linewidth]{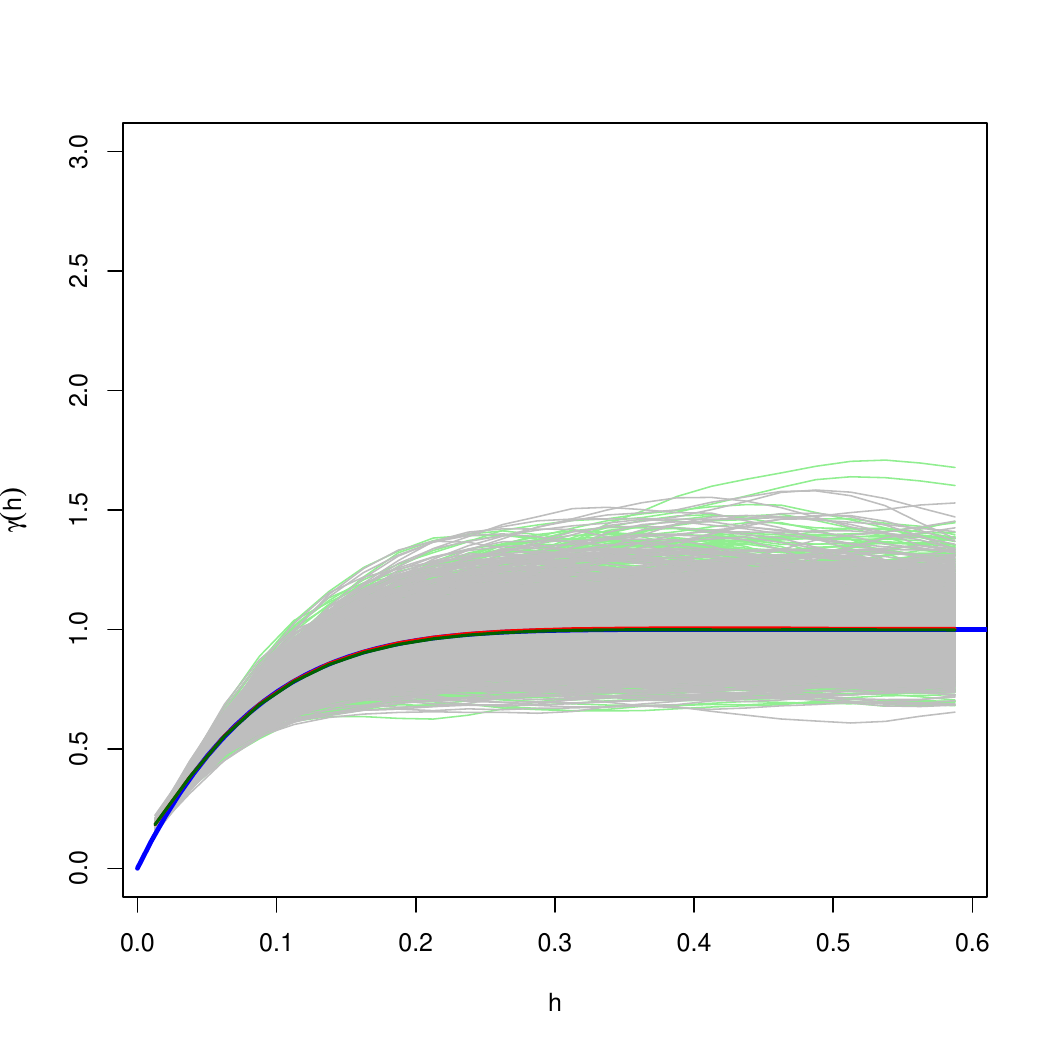}
        \caption{Scenario 7}
        \label{fig:7}
    \end{subfigure}
    \hfill
    \begin{subfigure}{0.45\textwidth}
        \centering
        \includegraphics[width=\linewidth]{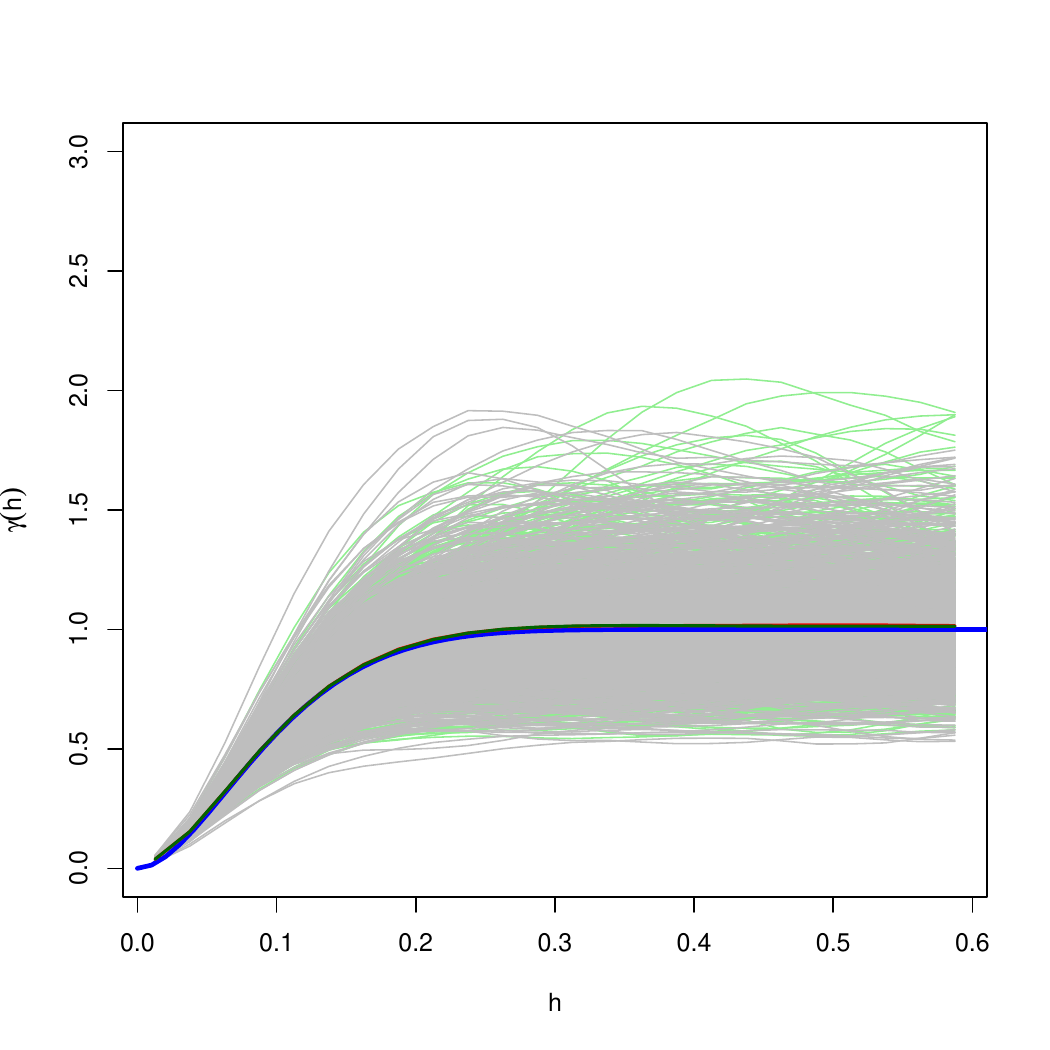}
        \caption{Scenario 8}
        \label{fig:8}
    \end{subfigure}

    \caption{Empirical semivariograms from $1{,}000$ realizations of a Gaussian random field under different parameter configurations of the $\mathcal{GW}_{\nu,\mu,a}$ model (see Table~\ref{tab:tabla2}), generated using the Gasper--mixture algorithm. Gray lines show the empirical semivariograms for individual realizations, red lines their empirical mean over the $1{,}000$ realizations, and blue lines the corresponding theoretical semivariograms. Green lines show empirical semivariograms computed from realizations obtained via the Cholesky method.}
    \label{fig:simu2}
\end{figure}

As a diagnostic of simulation quality, we follow \citet{emery2016,cuevas2019fast,arroyo2021} and compute the omnidirectional empirical semivariogram for each realization.
For each parameter configuration, the empirical semivariograms are compared with their theoretical counterparts.
Across the scenarios considered, the average empirical semivariograms track the theoretical ones closely, indicating that the proposed finite-$L$ STB schemes reproduce the target second-order structure well.
The results are shown in Figures~\ref{fig:simu1} and~\ref{fig:simu2}, where, for reference, we also report empirical semivariograms obtained from realizations generated by the Cholesky method.

The same experimental design is used for the Kummer--Tricomi model $\mathcal{K}_{\nu,\mu,\beta}$.
Since $\nu$ controls mean-square differentiability, we consider $\nu=0.5$ for non-differentiable realizations and $\nu=1.5$ for smoother realizations.
We also consider two scale parameters, $\beta=0.1$ and $\beta=0.5$, which are related to the correlation range.
The tail parameter $\mu$ governs the large-distance behavior of the correlation function and determines whether the model exhibits short-range or long-range dependence.
Here we consider $\mu=0.25$ (long-range dependence) and $\mu=3.5$ (short-range dependence), yielding the six scenarios summarized in Table~\ref{tab:tabla3}.

\begin{figure}[htbp]
    \centering
    \begin{subfigure}{0.35\textwidth}
        \centering
        \includegraphics[width=0.95\linewidth]{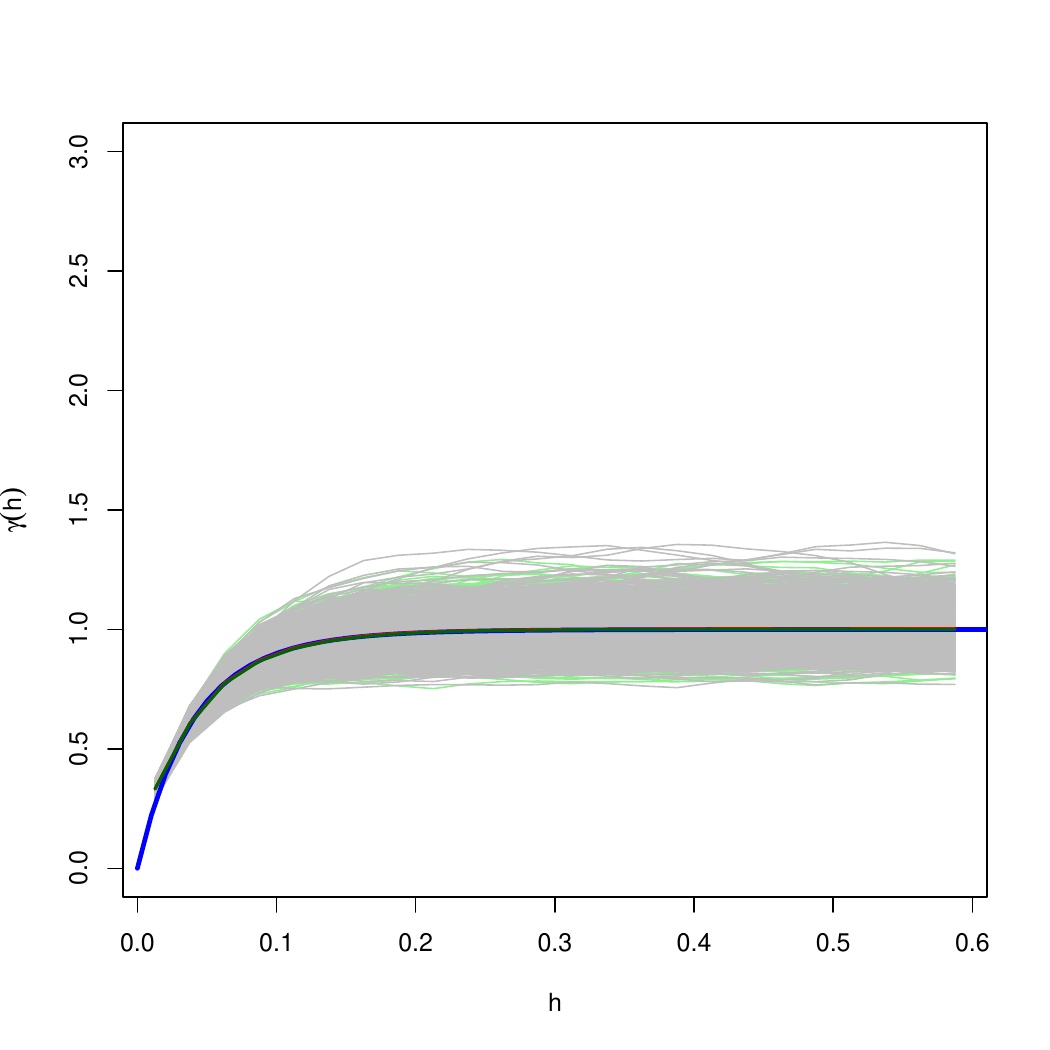}
        \caption{Scenario 9}
    \end{subfigure}
    \hfill
    \begin{subfigure}{0.35\textwidth}
        \centering
        \includegraphics[width=0.95\linewidth]{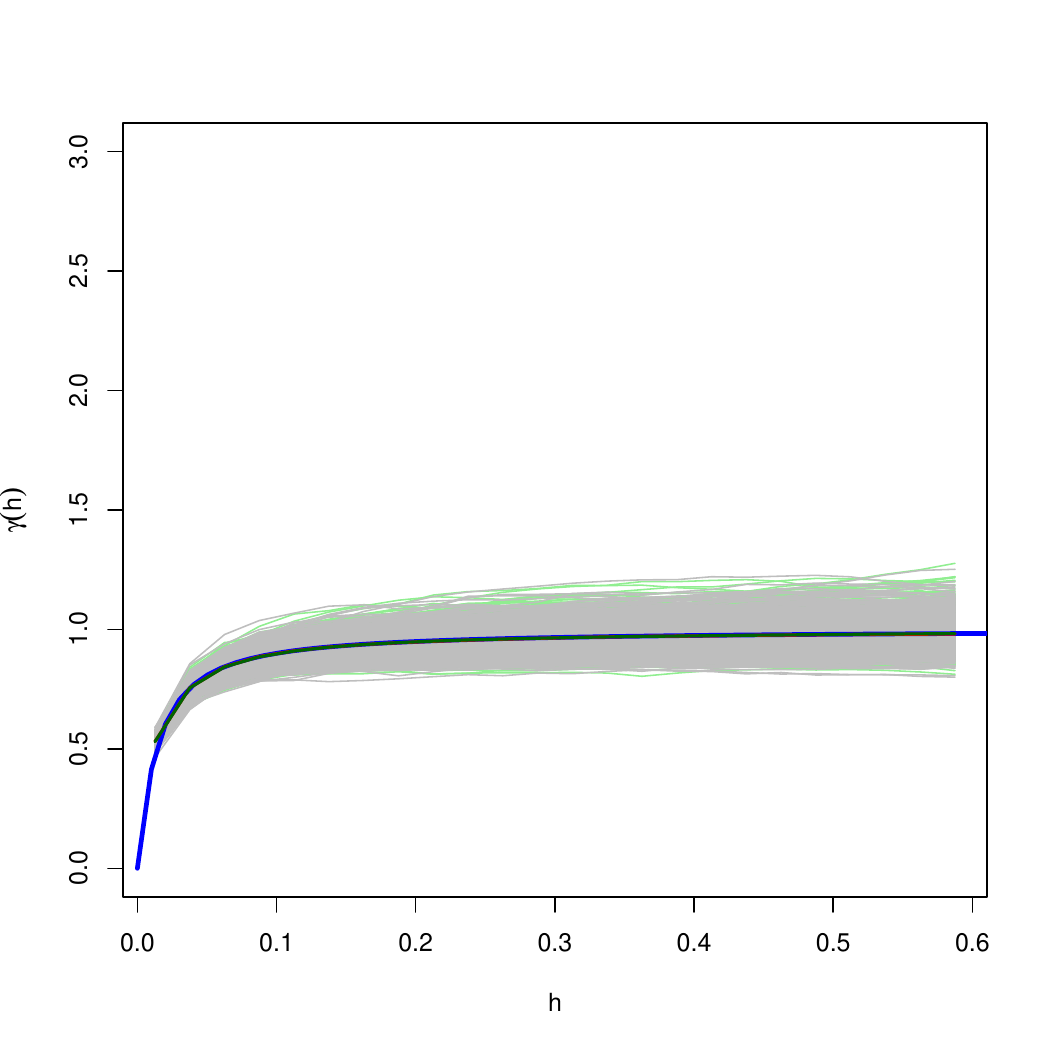}
        \caption{Scenario 10}
    \end{subfigure}
    \vspace{0.5cm}
    \begin{subfigure}{0.35\textwidth}
        \centering
        \includegraphics[width=0.95\linewidth]{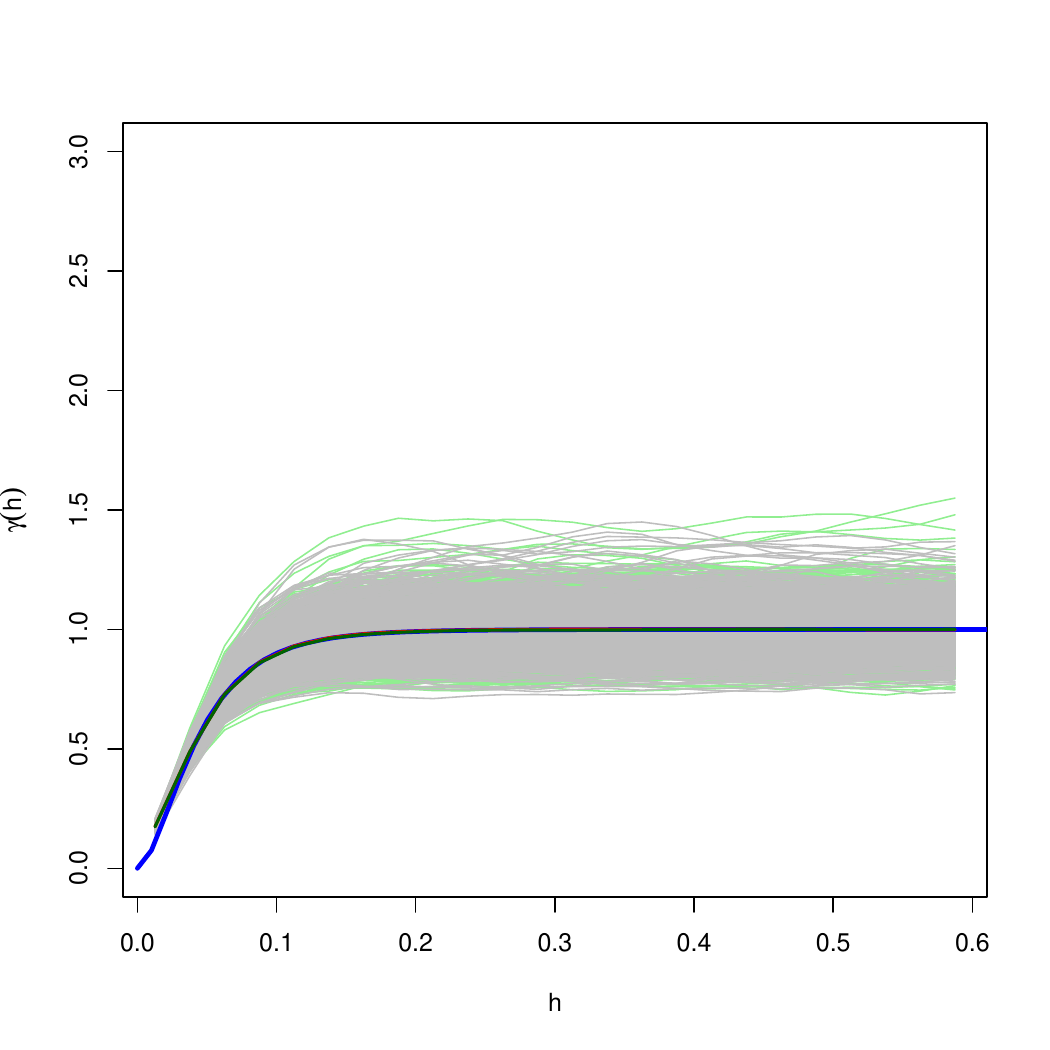}
        \caption{Scenario 11}
    \end{subfigure}
    \hfill
    \begin{subfigure}{0.35\textwidth}
        \centering
        \includegraphics[width=0.95\linewidth]{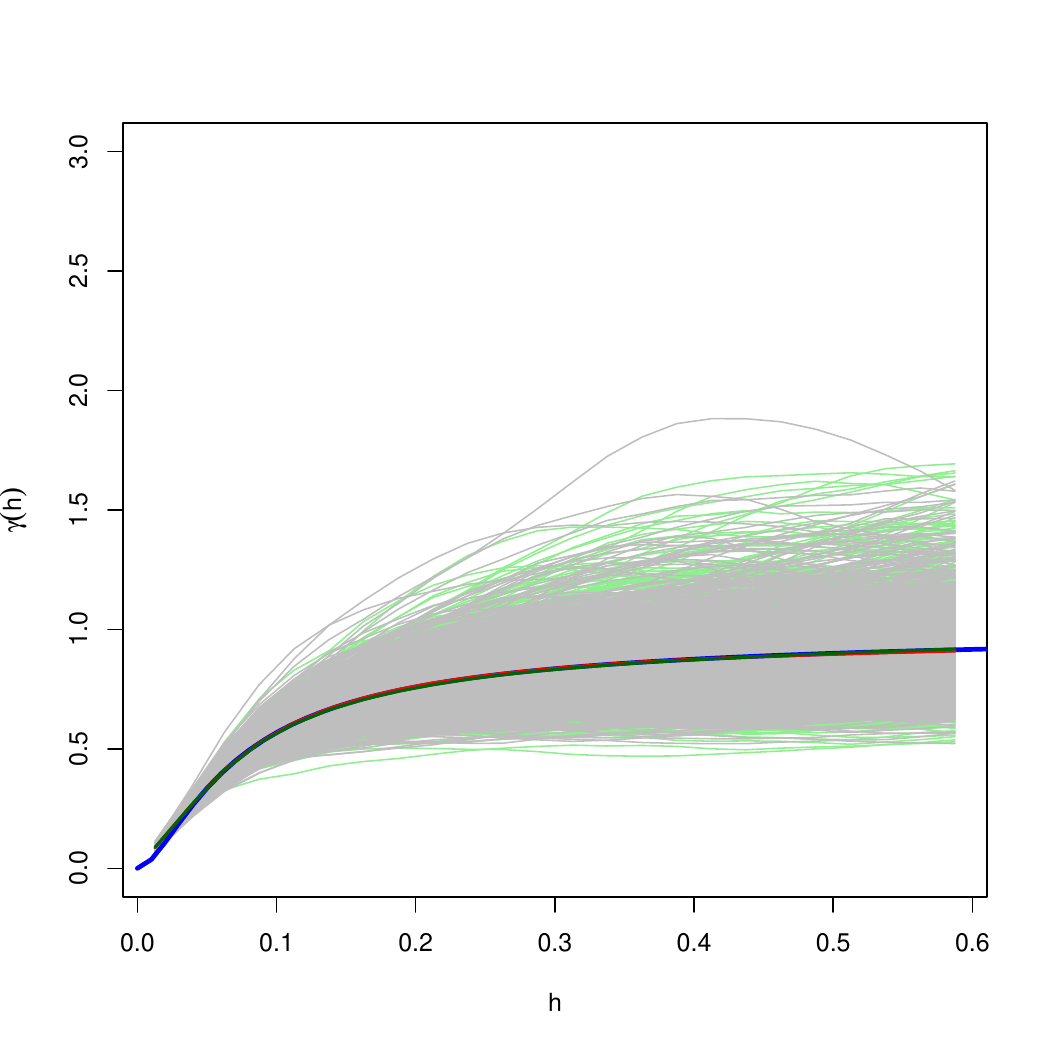}
        \caption{Scenario 12}
    \end{subfigure}
    \vspace{0.5cm}
    \begin{subfigure}{0.35\textwidth}
        \centering
        \includegraphics[width=0.95\linewidth]{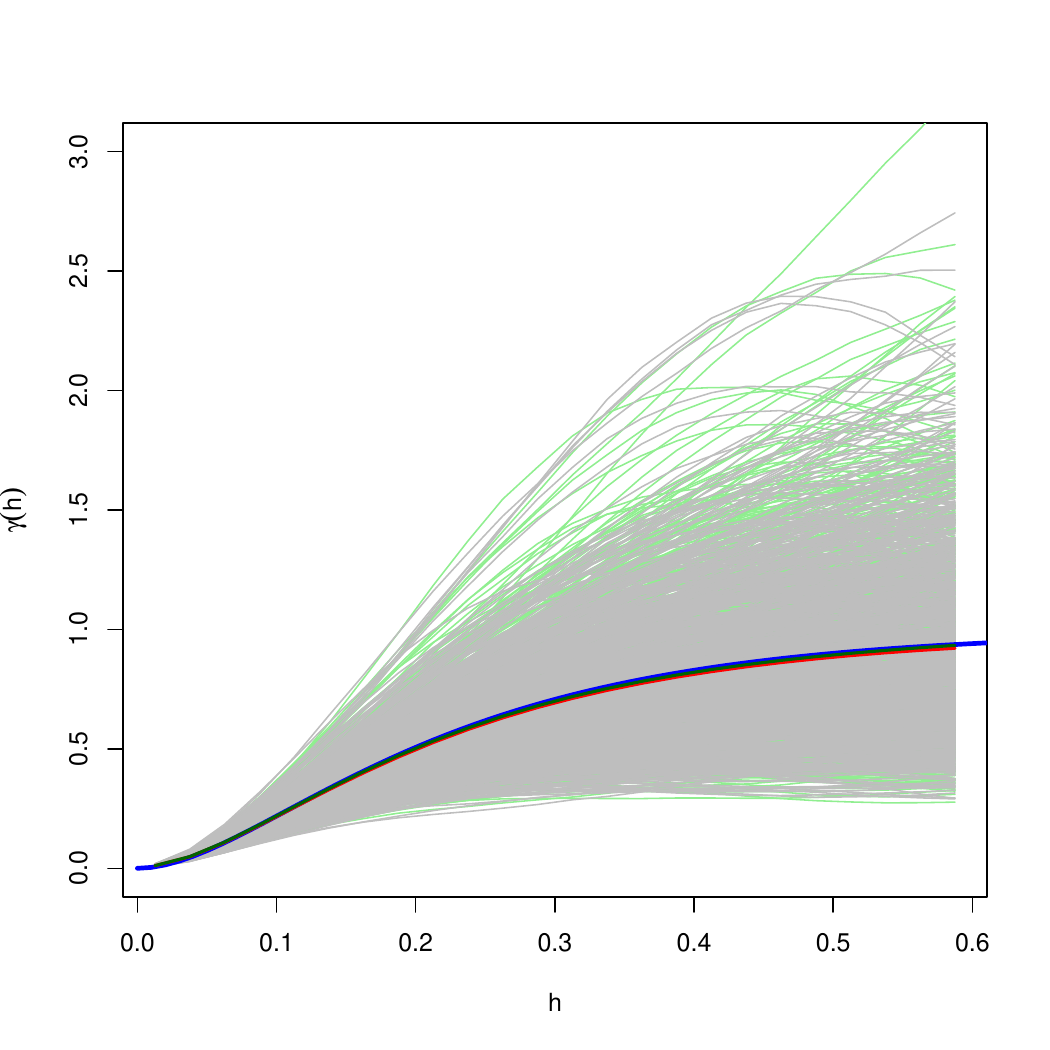}
        \caption{Scenario 13}
    \end{subfigure}
    \hfill
    \begin{subfigure}{0.35\textwidth}
        \centering
        \includegraphics[width=0.95\linewidth]{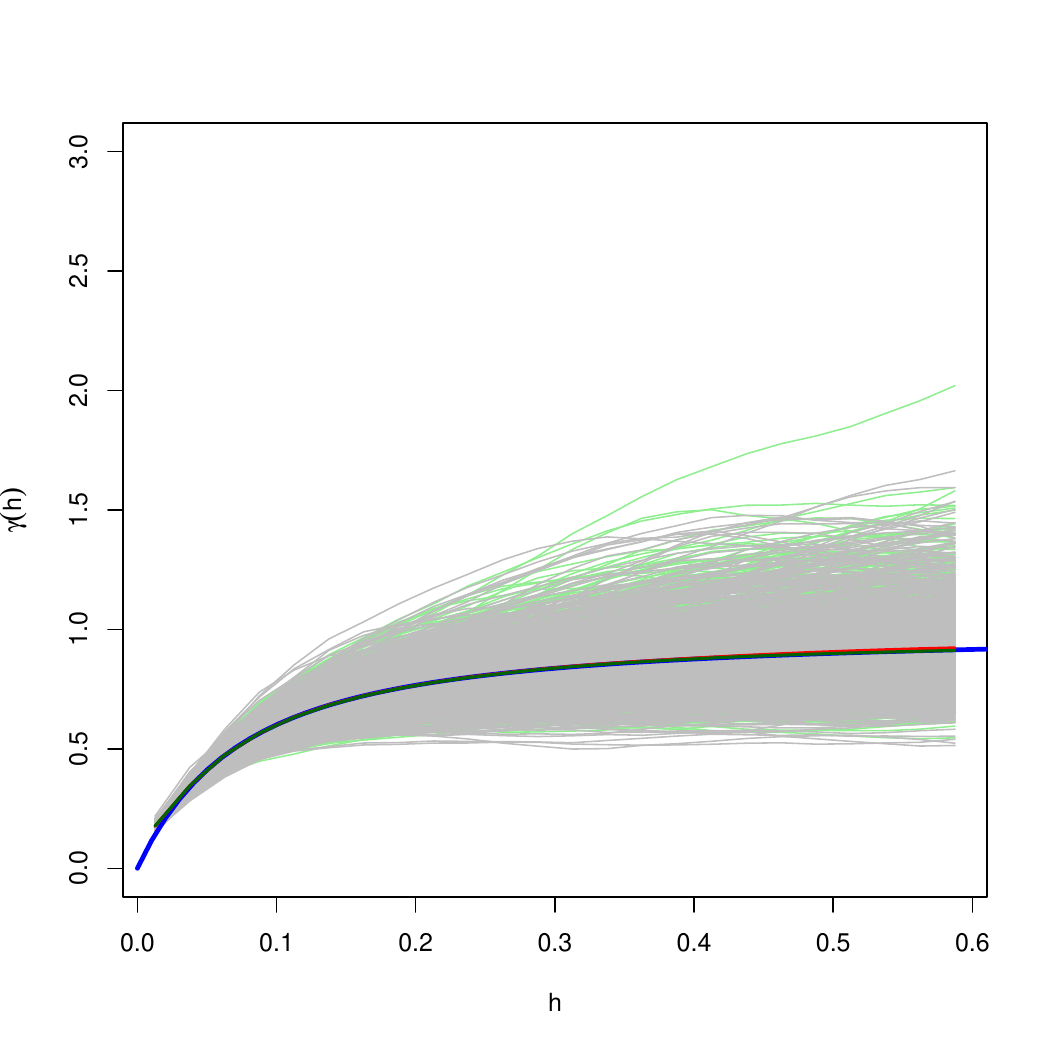}
        \caption{Scenario 14}
    \end{subfigure}
    \caption{Empirical semivariograms from $1{,}000$ realizations of a Gaussian random field under different parameter configurations of the $\mathcal{K}_{\nu,\mu,\beta}$ model (see Table~\ref{tab:tabla3}), generated using the proposed STB algorithm. Gray lines show the empirical semivariograms for individual realizations, red lines their empirical mean over the $1{,}000$ realizations, and blue lines the corresponding theoretical semivariograms. Green lines show empirical semivariograms computed from realizations obtained via the Cholesky method.}
    \label{fig:simu3}
\end{figure}

As in the generalized Wendland case, the empirical semivariograms agree closely with their theoretical counterparts across the parameter settings considered.
The results, reported in Figure~\ref{fig:simu3}, indicate that the proposed STB scheme reproduces the target semivariogram accurately for both short-range and long-range dependence.

\begin{table}[htbp]
\centering
\caption{
Parameters of the $\mathcal{GW}_{\nu,\mu,a}$ correlation model used in the simulation study. The parameters are chosen such that $\mu/2 - 1 - \nu > 0$, corresponding to the Beta--mixture algorithm (Section~5.1).}
\begin{tabular}{|c|c|c|c|}
\hline
\textbf{Scenario} & \textbf{Smoothness ($\nu$)} & \textbf{Compact support ($a$)} & \textbf{Shape ($\mu$)} \\
\hline
1 & 0 & 0.1 & 6 \\
\hline
2 & 1 & 0.1 & 7 \\
\hline
3 & 0 & 0.5 & 6 \\
\hline
4 & 1 & 0.5 & 7 \\
\hline
\end{tabular}
\label{tab:tabla1}
\end{table}

\begin{table}[htbp]
\centering
\caption{Parameters of the $\mathcal{GW}_{\nu,\mu,a}$ correlation model used in the simulation study. The parameters are chosen such that $\mu/2 - 1 - \nu < 0$, corresponding to the Gasper--mixture algorithm (Section~5.2).}
\begin{tabular}{|c|c|c|c|}
\hline
\textbf{Scenario} & \textbf{Smoothness ($\nu$)} & \textbf{Compact support ($a$)} & \textbf{Shape ($\mu$)} \\
\hline
5 & 0 & 0.1 & 2 \\
\hline
6 & 1 & 0.1 & 3 \\
\hline
7 & 0 & 0.5 & 2 \\
\hline
8 & 1 & 0.5 & 3 \\
\hline
\end{tabular}
\label{tab:tabla2}
\end{table}

\begin{table}[htbp]
\centering
\caption{Parameters of the $\mathcal{K}_{\nu,\mu,\beta}$ correlation model used in the simulation study. The tail parameter $\mu$ is chosen to explore short-range dependence (Scenarios~9, 11, 13) and long-range dependence (Scenarios~10, 12, 14), while the scale parameter $\beta$ is chosen such that $\mathcal{K}_{\nu,\mu,\beta}(0.1)=0.90$ (Scenarios~9--11) or $\mathcal{K}_{\nu,\mu,\beta}(0.5)=0.90$ (Scenarios~12--14).}
\begin{tabular}{|c|c|c|c|}
\hline
\textbf{Scenario} & \textbf{Smoothness ($\nu$)} & \textbf{Scale ($\beta$)} & \textbf{Tail ($\mu$)} \\
\hline
9  & 0.5 & 0.101 & 3.5 \\
\hline
10 & 0.5 & 0.013 & 0.25 \\
\hline
11 & 1.5 & 0.059 & 3.5 \\
\hline
12 & 1.5 & 0.032 & 0.25 \\
\hline
13 & 1.5 & 0.293 & 3.5 \\
\hline
14 & 0.5 & 0.064 & 0.25 \\
\hline
\end{tabular}
\label{tab:tabla3}
\end{table}

\subsection{Global Rank Envelope Test for additional distributional assessment}

To complement the semivariogram-based comparisons, we perform an additional diagnostic based on the \emph{Global Rank Envelope Test} (GET) of \citet{globale}, implemented in the \texttt{GET} \texttt{R} package.
The goal is not to establish full distributional equivalence of the simulated fields, but rather to assess whether the proposed STB algorithms reproduce the distribution of empirical semivariograms generated under the target model.

The experimental setup is as follows.
We consider three correlation models: the generalized Wendland model $\mathcal{GW}_{0,2,0.1}$, the Kummer--Tricomi model $\mathcal{K}_{0.5,3.5,0.076}$, and the Mat\'ern model $\mathcal{M}_{0.5,0.15}$, over $n=1{,}000$ spatial locations drawn uniformly on $[0,1]^2$.
For each model and for each of $1{,}000$ independent replications, we proceed as follows.
First, one realization is generated, separately by the Cholesky method and by the proposed STB algorithm, and its omnidirectional empirical semivariogram is computed using $25$ bins up to maximum distance $0.6$.
Second, an additional ensemble of $2{,}499$ independent STB realizations is generated, and the corresponding empirical semivariograms are used as the simulated reference set.
The GET is then applied by treating the semivariogram of the template realization as the observed functional statistic and the ensemble of $2{,}499$ semivariograms as the reference sample.
Under the null hypothesis that the template semivariogram is drawn from the same distribution as the reference ensemble, the resulting $p$-values should follow a $\mathrm{Unif}(0,1)$ distribution.

\begin{figure}[htbp]
    \centering
    \begin{subfigure}{0.32\textwidth}
        \centering
        \includegraphics[width=\linewidth]{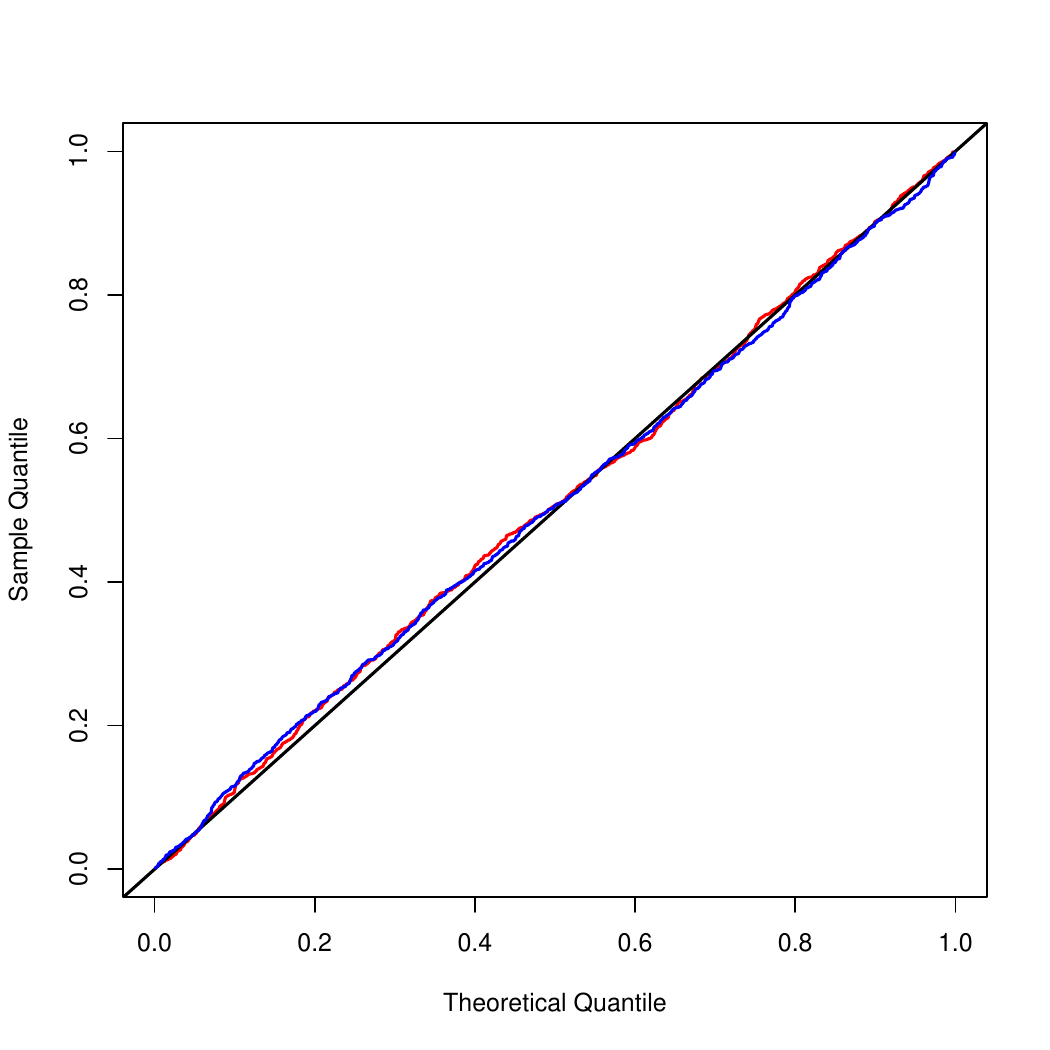}
        \caption{$\mathcal{GW}_{0,2,0.1}$}
        \label{fig:GET_GW}
    \end{subfigure}
    \hfill
    \begin{subfigure}{0.32\textwidth}
        \centering
        \includegraphics[width=\linewidth]{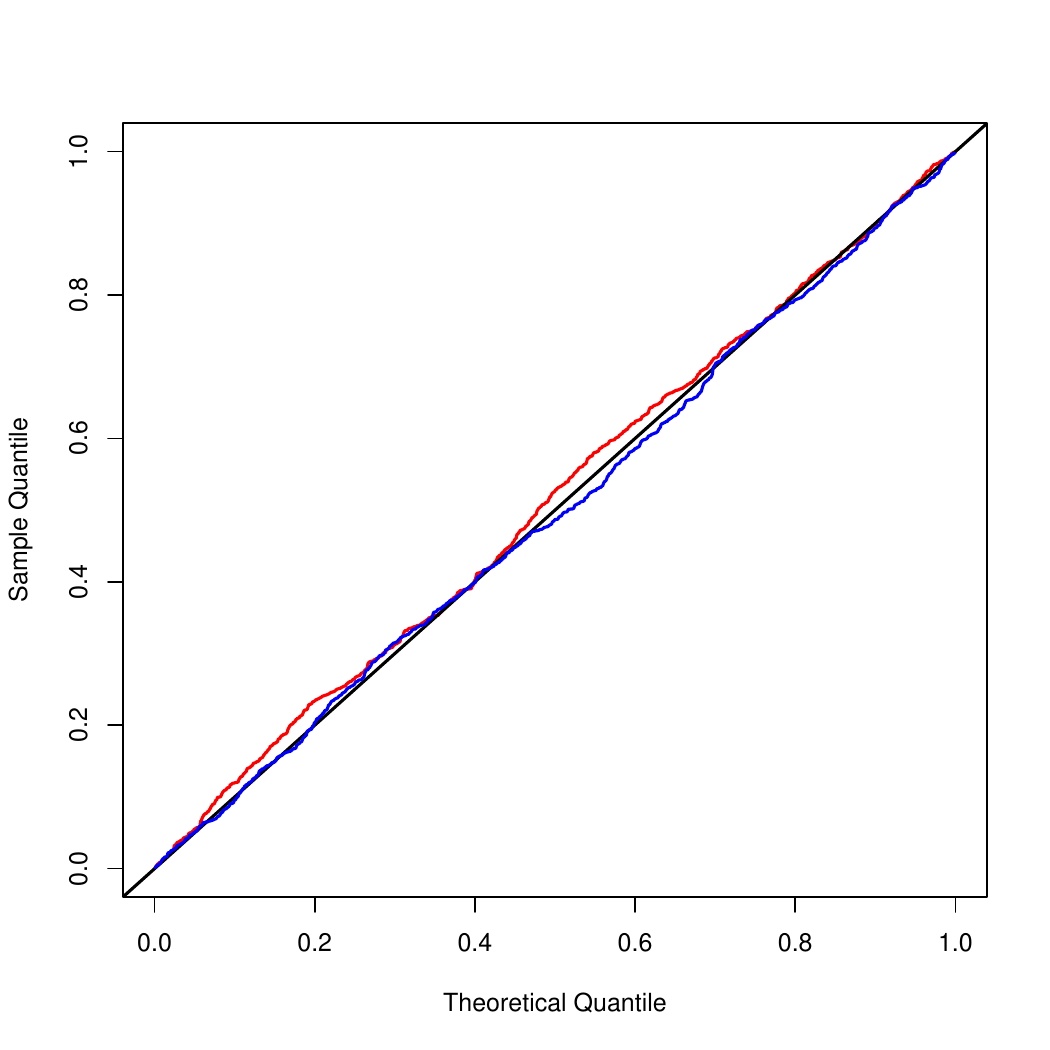}
        \caption{$\mathcal{K}_{0.5,3.5,0.076}$}
        \label{fig:GET_K}
    \end{subfigure}
    \hfill
    \begin{subfigure}{0.32\textwidth}
        \centering
        \includegraphics[width=\linewidth]{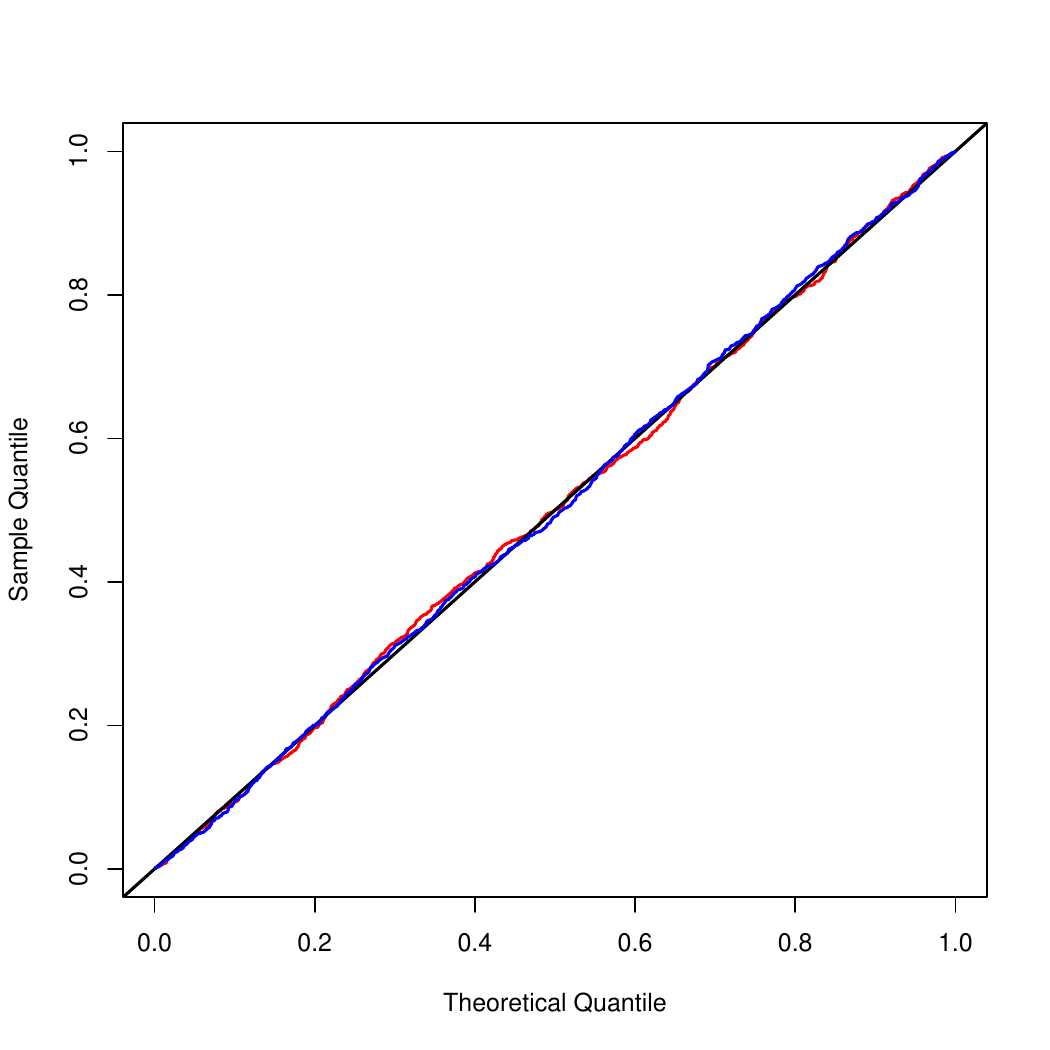}
        \caption{$\mathcal{M}_{0.5,0.15}$}
        \label{fig:GET_M}
    \end{subfigure}
    \caption{Quantile--quantile plots of empirical $p$-value distributions against $\mathrm{Unif}(0,1)$, based on $B=1{,}000$ replications of the Global Rank Envelope Test. Each panel corresponds to one correlation model. Red curves correspond to Cholesky-generated templates and blue curves to STB-generated templates. The diagonal line is the $45^\circ$ reference.}
    \label{fig:GET}
\end{figure}

The test is therefore applied twice per replication: once with a Cholesky-generated template and once with an STB-generated template.
Over the $1{,}000$ replications, this yields two empirical distributions of $p$-values for each model.
Figure~\ref{fig:GET} reports quantile--quantile plots of these empirical $p$-value distributions against the $\mathrm{Unif}(0,1)$ law.

For all three covariance models, the $p$-value distributions obtained from STB-generated templates are close to those obtained from Cholesky-generated templates and remain near the $45^\circ$ reference line.
This provides additional evidence that the proposed STB algorithms reproduce well the distribution of empirical semivariograms under the target model.
In particular, the diagnostic supports the conclusions drawn from the semivariogram comparisons for models with compact support ($\mathcal{GW}$), polynomially decaying tails ($\mathcal{K}$), and classical Mat\'ern dependence.

\subsection{Computational performance}

As a final numerical experiment, we assess how the proposed STB algorithms scale with the number of spatial locations in very large simulation settings. Table~\ref{tab:large-scale-times} reports the median elapsed time, computed over $n_{\rm sim}=10$ independent realizations, for $n=500{,}000$, $1{,}000{,}000$, $5{,}000{,}000$, $10{,}000{,}000$, and $50{,}000{,}000$ locations.

We consider three representative scenarios: Scenario~1 of the $\mathcal{GW}_{0,6,0.1}$ model, corresponding to the Beta--mixture algorithm; Scenario~6 of the $\mathcal{GW}_{1,3,0.1}$ model, corresponding to the Gasper--mixture algorithm; and Scenario~9 of the $\mathcal{K}_{0.5,3.5,0.101}$ model. The results show a nearly linear increase of computational time with the number of locations for all three cases, in agreement with the $O(NL)$ complexity of the STB construction when the number of spectral components is fixed. Moreover, the computational costs of the Beta--mixture and Gasper--mixture implementations are very similar in practice, indicating that the additional discrete mixture step in the Gasper case has only a limited impact on overall runtime. The $\mathcal{K}_{0.5,3.5,0.101}$ model exhibits essentially the same scaling behavior. Overall, these results confirm that the proposed methods remain computationally feasible even for simulations involving tens of millions of locations.

\begin{table}[htbp]
\centering
\caption{Median elapsed time (in seconds), computed over $n_{\rm sim}=10$ independent realizations, for large-scale STB simulations at increasing numbers of spatial locations. Scenario~1 corresponds to the $\mathcal{GW}_{0,6,0.1}$ model with the Beta--mixture algorithm, Scenario~6 to the $\mathcal{GW}_{1,3,0.1}$ model with the Gasper--mixture algorithm, and Scenario~9 to the $\mathcal{K}_{0.5,3.5,0.101}$.}
\label{tab:large-scale-times}
\begin{tabular}{lccccc}
\hline
Model / Scenario & $500{,}000$ & $1{,}000{,}000$ & $5{,}000{,}000$ & $10{,}000{,}000$ & $50{,}000{,}000$ \\
\hline
$\mathcal{GW}_{0,6,0.1}$, Scenario~1 (Beta--mixture) & 3.90 & 7.87 & 39.50 & 78.54 & 395.40 \\
$\mathcal{GW}_{1,3,0.1}$, Scenario~6 (Gasper--mixture) & 3.90 & 7.87 & 40.07 & 80.33 & 399.50 \\
$\mathcal{K}_{0.5,3.5,0.101}$, Scenario~9 & 3.85 & 7.85 & 39.27 & 78.43 &  394.51 \\
\hline
\end{tabular}
\end{table}

\section{Conclusions}
\label{sec8}

We have revisited the spectral turning--bands (STB) method as a general and scalable framework for simulating isotropic Gaussian random fields with flexible correlation structures. By exploiting probabilistic representations of the spectral measure, we derived exact sampling schemes for the spectral variables entering the STB construction not only for the Mat\'ern ${\cal M}$ model, but also for its Kummer--Tricomi ${\cal K}$ and Gauss--Hypergeometric ${\cal GH}$ extensions, which allow for heavy tails, long-range dependence, and compactly supported correlations, including Generalized Wendland models as special cases. The resulting procedures retain the usual finite-$L$ STB approximation to the target Gaussian random field, while leading to simple algorithms with linear computational cost in the number of locations and spectral components. These methods are implemented in the \texttt{GeoModels} \texttt{R} package, making them readily applicable to large-scale simulation studies and to the generation of non-Gaussian fields through transformations of latent Gaussian processes.

A natural direction for future work is the development of fast STB simulation methods for multivariate random fields with ${\cal GH}$ and ${\cal K}$ matrix-valued covariance models, where diagonal entries describe the spatial dependence within each component and off-diagonal entries characterize cross-dependence between components. A few multivariate constructions for the ${\cal GH}$ class have been proposed by \cite{daley2015} and \cite{emery2022gauss}, in which both direct and cross-covariances belong to the ${\cal GH}$ family under suitable parametric constraints. By contrast, to the best of our knowledge, no corresponding study is currently available for multivariate ${\cal K}$ models. Extending the univariate STB construction to the multivariate setting is nontrivial, since matrix-valued covariance functions are associated with matrix-valued spectral densities that cannot be sampled in the same way as in the univariate case; consequently, importance-sampling strategies or related techniques are often required \citep{emery2016}.

\section*{Acknowledgments}

This work was partially funded and supported by the National Agency for Research and Development of Chile, through grants ANID FONDECYT REGULAR 1240308 (M. Bevilacqua), ANID FONDECYT INICIACION 11240330 (F. Cuevas-Pacheco) and ANID PIA CIA250010 (X. Emery).
The authors confirm that the data and code  supporting the findings of this study are available in the supplementary materials.

\newpage

\bibliography{sample}


\end{document}